\documentclass[useAMS,usenatbib]{mn2e}
\usepackage{graphicx,amsmath,multirow,amssymb}
\usepackage{natbib}
\newcommand{\comment}[1]{}

\def\simgt{\lower.5ex\hbox{$\; \buildrel > \over \sim \;$}}
\def\simlt{\lower.5ex\hbox{$\; \buildrel < \over \sim \;$}}

\title{Dust from AGBs: relevant factors and modelling uncertainties}
\author[Ventura et al.]{P. Ventura$^1$, F. Dell'Agli$^{1,3}$, R. Schneider$^1$, 
M. Di Criscienzo$^{1,2}$, C. Rossi$^{3}$,   
\newauthor
F. La Franca$^{4}$, S. Gallerani$^5$, R. Valiante$^{1}$  \\
$^1$INAF -- Osservatorio Astronomico di Roma, Via Frascati 33, 00040, Monte Porzio Catone (RM), Italy \\
$^2$INAF -- Osservatorio Astronomico di Capodimonte, Salita Moiarello 16, 80131, Napoli, Italy \\
$^{3}$Dipartimento di Fisica, Universit\`a di Roma ``La Sapienza'', P.le Aldo Moro 5, 00143, 
Roma, Italy \\
$^{4}$Dipartimento di Matematica e Fisica, Universit\`a degli Studi ``Roma Tre'', Via della Vasca Navale 84, 00146,  Roma, Italy \\
$^{5}$Scuola Normale Superiore, Piazza dei Cavalieri 7, 56126 Pisa, Italy\\
}

\begin{document}

\date{Accepted, Received; in original form }

\pagerange{\pageref{firstpage}--\pageref{lastpage}} \pubyear{2012}

\maketitle

\label{firstpage}

\begin{abstract}
The dust formation process in the winds of Asymptotic Giant Branch stars is discussed, based
on full evolutionary models of stars with mass in the range 
$1$M$_{\odot} \leq$M$\leq 8$M$_{\odot}$, and metallicities 
$0.001 < Z <0.008$. Dust grains are assumed to form in an isotropically expanding wind,
by growth of pre--existing seed nuclei.

Convection, for what concerns the treatment of convective borders and the efficiency of
the schematization adopted, turns out to be the physical ingredient used to calculate the
evolutionary sequences with the highest impact on the results obtained.

Low--mass stars with M$\leq 3$M$_{\odot}$ produce carbon type dust with also
traces of silicon carbide. The mass of solid carbon formed, fairly independently of 
metallicity, ranges from a few $10^{-4}$M$_{\odot}$, for stars of initial mass 
$1-1.5$M$_{\odot}$, to $\sim 10^{-2}$M$_{\odot}$ for M$\sim 2-2.5$M$_{\odot}$; 
the size of dust particles is in the range $0.1 \mu$m$\leq a_C \leq 0.2\mu$m. 
On the contrary, the production of silicon carbide (SiC) depends on metallicity. 
For $10^{-3} \leq Z \leq 8\times 10^{-3}$ the size of SiC grains varies in the
range $0.05 \mu {\rm m} <  {\rm a_{SiC}} < 0.1 \mu$m, while the mass of SiC formed is 
$10^{-5}{\rm M}_{\odot} < {\rm M_{SiC}} < 10^{-3}{\rm M}_{\odot}$.

Models of higher mass experience Hot Bottom Burning, which prevents the formation
of carbon stars, and favours the formation of silicates and corundum. In this case the
results scale with metallicity, owing to the larger silicon and aluminium contained in
higher--Z models. At Z=$8\times 10^{-3}$ we find that the most massive stars produce
dust masses $m_d \sim 0.01$M$_{\odot}$, whereas models of smaller mass produce
a dust mass ten times smaller. The main component of dust are silicates, although 
corundum is also formed, in not negligible quantities ($\sim 10-20\%$).
 
\end{abstract}


\section{Introduction}
During the last years the evolution experienced by stars of intermediate mass 
after the consumption of central helium has received a growing interest by the 
astrophysical community. This phase, known as Asymptotic Giant Branch 
\citep[AGB,][]{iben83, lattanzio87, herwig05}, is much
shorter than the previous stages of core hydrogen and helium burning, but is
extremely important for the role played by these sources in the 
pollution of the interstellar medium. During the AGB evolution the stars loose all
their external mantle, returning into the interstellar medium material whose
chemical composition was altered by internal nuclear processes.

AGBs have been suggested as the main contributors to the pollution of the interstellar
medium in Globular Clusters, giving rise to the formation of multiple 
populations \citep{ventura01}.
The gas ejected by AGBs, possibly diluted with pristine gas, may stimulate the formation 
of one or more additional stellar components, overlapped to the original population present 
in the cluster \citep{dercole08, dercole10}. 
This scenario can explain most of the observational data, such as the chemical
anomalies involving light elements \citep{gratton12} and the photometric features of the main 
sequence and of the horizontal branch of Globular Clusters \citep{piotto09}.

The interest towards AGBs stems also from the physical conditions of their circumstellar
envelopes, particularly suitable to gas condensation into dust grains. The surface layers 
of these stars are sufficiently cool (${\rm T_{eff}} < 4000$K) to allow dust formation at 
typical distances of 3--10 stellar radii from the surface, where the densities are still 
sufficiently large \citep[$\rho > 10^{-14} \rm{gr/cm^3}$,][]{gs85} to allow formation of 
meaningful quantities of dust. 

Dust formation is a complex phenomenon, which is strongly interfaced with the dynamical
and thermal structure of the winds of this class of objects \citep{bert85, wood79, bowen88}. 
Stellar pulsation is known to trigger the formation of periodic shocks, that favours
the increase in the local density, thus stimulating gas condensation in cool regions in 
the stellar surroundings. Radiation pressure on dust grains is currently believed as the 
dominant mechanism triggering mass loss in these structures \citep{fleischer92}.

A self--consistent description of the dust formation process would require full coupling of
results from stellar evolution modelling with the description of the wind. A natural
outcome of such a treatment would be the outwards velocity of gas and dust particles 
leaving the star, thus the mass loss rate experienced. Because we are still far from 
achieving this target, the few models available in the literature treat the wind 
independently, based on the results coming from the integration of the equations of stellar 
equilibrium, from which the values of luminosity, effective temperature and mass 
loss rate of the star are obtained \citep{fg01, fg02, fg06}. These studies may be considered 
as a preliminary step towards a more complete treatment, where central object and external 
wind are described simultaneously. 

The reliability of the results obtained, in terms of the amount of dust formed, 
is hampered by the uncertainties associated to the wind model, for what concerns, 
e.g., the initial velocity of gas particles, the opacities relevant to determine the 
radiation pressure on grains, the sticking coefficients of each dust species 
\citep{fg06, paperI, paperII}.

AGB modelling is also relevant for the results obtained. The type of 
grains formed, either silicates or carbonaceous particles, depends
on the description of the processes able to alter the surface chemistry of AGBs, i.e. 
Third Dredge--Up (TDU) and Hot Bottom Burning (HBB). The efficiency of either mechanisms
is affected by the modelling of convection and in particular by the treatment
of the convective/radiative interface \citep{herwig00, herwig05} and of the temperature 
gradient in regions unstable to convective motions \citep{vd05a}.
The description of mass loss is also extremely relevant, because it has
a direct impact on the AGB evolutionary time scale \citep{vd05b}; furthermore, the mass loss 
rate determines the density profile of the wind, thus the amount of dust formed.

It is for all these reasons that results in the literature presented by different
research groups, though using the same description of the wind and of the dust grains 
formation and growth, differ considerably in the mass and composition of dust formed
\citep{fg06, paperI, paperII, dicrisci13, nanni13}.

The scope of this work is to discuss how the properties of dust
formed in the winds of AGBs depend on the implementation of some fundamental features
of the model, namely: i) the macrophysics adopted to describe the
AGB evolution; ii) the thermodynamical description of the wind: iii) the formation
and growth of grains. 

We present a new grid of dust formation models assuming stellar metallicities of
Z=$4\times 10^{-3}$, which fills the gap among the 
chemistries of our previous explorations, i.e. Z=$10^{-3}$ \citep{paperI}, 
Z=$8\times 10^{-3}$ \citep{paperII}, Z=$3\times 10^{-4}$ \citep{dicrisci13}. 
This allows to explore in details the role played by the chemical composition in the dust
formation process. Also, this metallicity will enable to compare model predictions
with observational data in the Small Magellanic Cloud \citep{larsen00} and in
metal--rich galactic Globular Clusters \citep{carretta09}.

In addition to our previous investigations on this topic, we present AGB models of
different metallicities where i) some extra--mixing is assumed from any convective border
(including the base of the external mantle); ii) the description of mass loss is based
on hydrodynamical wind models including carbon--dust formation \citep{wachter08}.

The paper is organized as follows: in section 2 we present an overview of the status of
the art of the modelling of dust formation around AGBs; sections 3 and 4 describe the inputs 
used to model the AGB evolution of stars and the dust grain growth in the expanding winds;
carbon dust formation is discussed in section 5, whereas silicates production in more
massive AGBs experiencing HBB is presented in section 6; finally, section 7 is devoted to 
an overall discussion of the results obtained.

\section{Dust formation modelling around AGBs: status of the art}
The pioneering explorations by the Heidelberg group \citep{fg01, fg02, fg06} set the 
framework to describe the dust formation process in the winds of AGBs. The wind is 
assumed to expand isotropically from the surface of the star, moving with 
velocities of a few km/s, entering into the dust formation layer, where gas (and dust) 
particles are accelerated via radiation pressure on the newly formed grains.

The dust grains are assumed to grow owing to the gas molecules impinging on the
already formed solid particles. The rate at which the grains grow scales with the
density of the gas; this defines a very narrow region where dust formation occurs.
An asymptotic behaviour is reached, with the wind expanding
outwards with large velocities (tipically $\sim 10-50$ km/s).

This is a simplified view of a much more complex situation, where large amplitude
pulsations trigger periodic shocks \citep{wood79, bowen88, 
bert85}: gas particles, rather than expanding at constant velocity, are projected on 
ballistic trajectories, where they contribute to the increase in the local density in 
regions very cool, thus extremely suitable to dust formation \citep{fleischer92}. 
A complete, self--consistent treatment would require the simultaneous description of 
the periodic variation of the stellar properties and of the structure of the wind, such 
that condensation takes place in the regions where the shock favours the increase 
in the density. A natural outcome of such a treatment would be the determination of 
the mass loss rate, via the density and the velocities of the gas leaving the star
\citep{wachter02, wachter08, mattsson08}.

Such a self--consistent approach is still missing and the models currently available
treat the wind independently, assuming as inputs the physical parameters of the
central star, i.e. effective temperature, luminosity, mass loss rate and the
surface chemical composition. The most relevant shortcoming
of this approach is the lack of any feedback between the formation of dust grains
and the mass loss rate, which is assumed apriori; we will come back to this point
in the following.

Because the structure of the wind is based on the properties of the central star, 
the effective temperature (T$_{\rm eff}$), luminosity (L) and mass loss 
rate ($\dot{\rm M}$) are key quantities in determining the amount of dust formed. 
Condensation of 
gas molecules is favoured at low T$_{\rm eff}$'s, because smaller temperatures in the 
wind allows the dust forming region to be closer to the surface of the star, in higher 
density zones. Larger values of $\dot{\rm M}$ favour dust formation, because mass 
conservation requires the density in the wind to increase linearly with $\dot{\rm M}$
(see Eq. \ref{eqro} below). Luminosity has no direct impact on the dust formation 
process, but it affects the acceleration due to radiation pressure, which scales with L
(Eq. \ref{eqv} and \ref{eqgamma} below).

The surface chemistry of the star is relevant in determining which kind of dust is formed.
The CO molecule has an extremely large dissociation energy (11eV), which means that the least
abundant between carbon and oxygen is bound in CO molecules. In C--rich environments there
is no oxygen available, thus the only dust that can be formed is solid carbon, 
silicon carbide (SiC) and iron \citep{groenewegen98}. When the C/O ratio is below unity 
all the carbon present is locked into CO: silicates, iron and corundum can form 
\citep{ossenkopf92}.

The physical evolution of AGBs, and of their surface chemistry, depends on some fundamental
physical properties such as convection and mass loss \citep{herwig05, karakas11}.

The treatment of the convective instability is the largest uncertainty 
affecting stellar evolution models. Owing to the lack of a self--consistent
theory of turbulence, the transport of energy within regions unstable to convective
motions is treated parametrically. The mixing length $\Lambda$, the typical scale of
convective motions, is assumed to be proportional to the pressure scale height H$_p$:
$\Lambda=\alpha$H$_p$. The parameter $\alpha$ is calibrated to reproduce the current radius 
of the Sun; with the latest updates in the micro--physics adopted, recent investigations
give $\alpha_{\odot}=1.75$ \citep{bressan12}.

The AGB phase is
the most sensitive to the convection model adopted. Models based on the Full Spectrum of 
Turbulence \citep[FST,][]{cm91} evolve at larger luminosities, on more expanded 
configurations, in comparison with their counterparts calculated with the traditional Mixing 
Length Theory \citep{vd05a}. FST models loose their envelopes more rapidly, and experience 
on the average a smaller number of thermal pulses (TP). 

The efficiency of the convective model adopted
has also an important feedback on the Hot Bottom Burning phenomenon, in models
with mass M$\geq3-3.5$M$_{\odot}$: the base of the convective envelope becomes sufficiently
hot (T$\geq 40$ MK) to favour an advanced proton capture nucleosynthesis, with the destruction 
of the surface carbon, and, when the temperatures exceed $\sim 70$ MK, the depletion of oxygen 
and the production of Aluminium via activation of the Mg--Al chain. HBB is found in all
models with core mass above $\sim 0.8$M$_{\odot}$ when convection is modelled according to
the FST formulation. Also MLT models with $\alpha > \alpha_{\odot}$ achieve much 
more easily HBB conditions \citep{renzini81, boothroyd88a}. The ignition of HBB is accompanied 
by the increase in the luminosity \citep{blocker91}, which, in turn, triggers the increase 
in the mass loss rate.

In models experiencing HBB the production of
silicates is greatly favoured: the C--star stage is inhibited by the destruction
of the surface carbon, and the star evolves at large luminosities, loose mass at a high rate,
which increase the dust formation rate. 

Low--mass AGBs (M below $3$M$_{\odot}$) do not reach HBB conditions, independently of 
convection modelling. In these stars the only mechanism able to alter the surface
chemistry is the Third Dredge--Up, which gradually enriches the surface in carbon,
eventually making C/O to exceed unity. Initially, when the mass loss rate is small and
the C--star stage is not yet reached, small quantities of silicates are produced;
in the final evolutionary stages carbon dust is produced in great quantities, owing to
the large surface carbon abundances and the high mass loss rates experienced.

While there is a general consensus on the sequence of events given above, the amount of
carbon dust formed is still matter of debate, because the evolutionary stage at
which the star becomes a C--star and the extent of the surface carbon enrichment depend
on the inwards penetration of the convective envelope during the TDU 
\citep[see discussion in][]{karakas11}. 
This is rendered uncertain by the poor knowledge of the behaviour of convective eddies
near the convective/radiative interface, particularly of the extent of the extra--mixing,
i.e. the distance they travel within radiatively stable regions.
The extra--mixing from convective cores of stars
in central burning phases can be calibrated via comparison with the observed main sequences 
of open clusters \citep{vandenbergh06}, by analysis of binary star data \citep{claret07},
and asterosismological investigations \citep{briquet07, montalban13}.
The situation for the AGB phase is more complex, as there is no observable that allows a
straight calibration of the overshoot from the base of the envelope, and from the
boundaries of the shell which forms at the ignition of each TP. 
The most widely argument used so far is the reproduction of the carbon star luminosity 
function in the Magellanic Clouds, which allows to determine the core mass at which TDU 
begins, and the extent of the inwards penetration of the convective envelope 
\citep{izzard04, marigo07}. These information can be used to estimate the extension of 
the mixed regions within the AGBs interiors.
The larger is the extra--mixing, the more carbon dust is produced, because more and more 
carbon is dredged--up to the surface regions.  

The description of the mass loss mechanism is also extremely relevant in determining the
evolution of AGBs \citep{herwig05}. The rate at which AGBs loose mass determines the
duration of the whole evolutionary phase \citep{vd05b}, but has also important consequences on the
modification of the surface chemistry. In massive AGBs, experiencing HBB, the mass
loss rate is stricly correlated with the degree of p--capture nuclesynthesis at the
bottom of the surface envelope: lower rates allow a stronger nucleosynthesis, thus
larger changes in the abundances of the elements involved in the CNO, Ne--Na and
Mg--Al cycles. In the low--mass regime, large rates favour faster consumption of the
envelope: the star undergoes a smaller number of TPs, thus the carbon surface 
enrichment is reduced.

This has a strong impact on the dust formation in the circumstellar envelopes. A 
direct effect of the mass loss description is associated with the higher density of the 
winds when the mass loss is large, which enhance dust production. Also, as stated above,
mass loss changes the surface chemical composition, and the mass fractions of those elements
relevant to allow dust formation (e.g carbon for carbon--type dust): this clearly 
changes the amount of dust which can be formed.

\section{Stellar evolution modelling}
The evolutionary sequences presented here were computed by means of the ATON
code for stellar evolution. The interested reader may find in \citet{ventura98} a
detailed description of the numerical structure of the code. The most recent updates can be 
found, e.g., in \citet{ventura09}. We recall here the main physical inputs, most relevant
for the topic of the present investigation.

\subsection{Convection}
The arguments presented in the previous section outline how relevant the description of
convection is for the AGB phase, in terms of the evolution of the main physical
quantities, number of thermal pulses experienced, change in the surface chemical composition
due to HBB and TDU.

In agreement with other investigations by our group, the temperature gradient within
convective regions is found by the FST model, developed by \citet{cm91}. 

Mixing of chemicals and nuclear burning were coupled with a diffusive approach, according
to the scheme suggested by \citet{cloutman}. The overshoot of convective eddies into
radiatively stable regions is described by an exponential decay of the velocity $\rm{v}$ 
beyond the formal border found via the Schwarzschild criterion. The relation used is

\begin{equation}
\rm{v}=\rm{v_b} \times \rm{exp}\left(\pm {1\over \zeta} \log{P\over P_b} \right)
\end{equation}

where the argument of the exponential is taken positive for overshoot from the upper
border of the convective zone, whereas the minus sign is used for overshoot below the
convective region. $\rm v_b$ and $\rm P_b$ are the values of velocity and pressure at the 
formal boundary. The extent of the overshoot is given by the free parameter $\zeta$.

For the convective cores that form during the hydrogen and helium burning phases, and generally
for the phases previous the AGB evolution, we followed the calibration given in \citet{ventura98}, 
and assumed $\zeta=0.02$.

For what concerns the AGB phase, we calibrated the extent of the extra--mixing from the 
base of the convective envelope and from the borders of the shell that forms at the
ignition of each thermal pulse based on the luminosity function (LF) of carbon stars in 
the Large Magellanic Cloud \citep{groenewegen04}. 
We find that adopting $\zeta=0.002$ leads to a satisfactory 
agreement between the observed and the predicted LFs. With this assumption we recover the
relation between initial mass and luminosity of the transition from O--rich to C--rich
surface chemical composition given in \citet{marigo07}.

\subsection{Mass loss}
In agreement with our previous explorations focused on the AGB evolution, we use 
for the phases preceeding the C--star stage the formulation by \citet{blocker95}. 
The Bl\"ocker formula ($\dot M \propto M^{-3.1}\times R \times L^{3.7}$, where M, R and
L are, respectively, the stellar mass, radius and luminosity) is suitable 
to describe the steep increase of $\dot M$ with luminosity
that characterizes massive AGBs; however, it underestimates the mass loss suffered by 
C--stars, because it is based on a description of the circumstellar envelope of MIRA
variables that neglects the carbon--dust formation process, and the radiation pressure
on dust grains, which induces a great acceleration of the wind. To describe the
C--star phase we rely on the formulation by \citet{wachter08}, giving the mass loss
rates experienced by low--mass AGBs as a function of mass, luminosity and effective
temperature. The expression used is 
$\dot M \propto M^{\alpha}T_{eff}^{\beta}L^{\gamma}$, where typical values for
the exponents (partly depending on the metallicity) are
$\alpha \sim -3$, $\beta \sim -7$, $\gamma \sim +3$.

\subsection{Radiative opacities}
The radiative opacities $k$ for temperatures above $10^4$~K were calculated using the OPAL 
online tool \citep{opal}; for smaller temperatures we used the AESOPUS tool described 
in \citet{marigo09}. This choice allows to account for the increase in the opacity 
associated with the change in the surface chemistry determined by TDU. The pioneering
study by \citet{marigo02} shows how the modelling of AGBs experiencing TDU is 
sensitive to the set of opacities adopted; detailed discussions on this argument can
be found in \citet{vm09, vm10}.

\subsection{Chemical composition}
The new models presented in this work have metallicity Z=0.004, initial helium
Y$=0.26$; the mixture is alpha--enhanced, with $[\alpha/$Fe$]=+0.2$. A discussion of 
their main physical properties, and of the change in the surface chemistry determined
by TDU and HBB, can be found in \citet{ventura13b}. Here we focus on the dust 
formed in their surroundings.

In addition to this new grid, we also present updated models for stellar
metallicities Z=0.001, Z=0.008 (discussed in previous papers) obtained with the
new implementation of the extra--mixing from convective regions during the AGB
phase and the mass loss description of the C--star phase.

\section{Dust grains growth}
The structure of the wind and the dust formation process is described following the
schematization by \citet{fg06}.
As discussed in the introduction, the thermodynamic structure of the wind and the
dust formation process are entirely determined by the physical and chemical conditions
at the surface of the central star. The latter is assumed to evolve independently of
the properties of the wind: this is possible because the mass loss rate is
determined via a parametric recipe, rather than by full coupling of star's and
wind's descriptions.

\subsection{Thermodynamic structure of the wind}
The wind is assumed to expand isotropically from the surface of the star, described by
a mass M, luminosity L, radius R, and effective temperature $T_{\rm eff}$. The thermodynamic structure
of the wind is set by the relations giving the variation of density ($\rho$) and
temperature (T) with the distance $r$ from the center of the star.

Indicating with $\dot M$ the rate of mass loss, we have

\begin{equation}
\dot M=4\pi r^2 \rho {\rm v},
\label{eqro}
\end{equation}
where v is the velocity of the wind.

The temperature of the gas in the wind decreases with $r$  according to the relation given 
by \citet{lucy76}

\begin{equation}
T={1\over 2}T_{\rm eff}^4 \left[ 1-\sqrt{1-{R^2\over r^2}}+{3\over 2}\tau \right].
\label{eqT}
\end{equation}

\noindent
The optical depth entering Eq.\ref{eqT} is found by the differential equation

\begin{equation}
{d\tau\over dr}=-\rho k {R^2\over r^2}.
\label{eqtau}
\end{equation}

\noindent
The boundary condition for Eq.\ref{eqtau} is that $\tau$ vanishes at infinity
(for numerical reasons, we impose that this condition is reached at a distance of
$10^4$R from the centre, where the asymptotic behaviour of all the quantities is
definitively reached).

The model is completed by the description of the acceleration of dust particles
under the effects of gravitational attraction from the star and radiation pressure
on dust grains:

\begin{equation}
{d \rm{v} \over dr}=-{GM\over r^2}\times (1-\Gamma),
\label{eqv}
\end{equation}

\noindent
with

\begin{equation}
\Gamma={kL\over 4\pi cGM}.
\label{eqgamma}
\end{equation}

For the wind velocity, we assume that it enters the dust forming region with a constant
velocity $\rm{v}_0=1$ km/sec, and is accelerated by radiation pressure. When dust formation
is efficient, the results show only a modest dependence on the assumed value for
$\rm{v}_0$.

\subsection{Dust grains formation and growth}
The system of equations given above can be closed if we know the opacity coefficient $k$,
needed to find $\Gamma$ (Eq.\ref{eqgamma}), and the optical depth (Eq.\ref{eqtau}).
$k$ can be found by summing the gas and dust contributions. The former is calculated by
means of the AESOPUS tool, whereas for the latter we calculate the individual contributions
from the various dust species formed.

In a carbon--rich context the species of dust formed are solid carbon,
iron and silicon carbide, whereas for M stars we consider
the formation of silicates (olivine, pyroxene and quartz), solid iron and corundum.
This latter species was not present in our previous investigations.

For each species it is defined a key--element, the least abundant among the elements present
in the molecules concurring to the condensation process (see Table 1 for the details of
the formation reactions of the various dust species considered). The abundance of the 
key--element
in the wind is extremely relevant for the amount of dust of a given species that can be
formed, because the growth rate ($J_i^{\rm gr}$) of the grains scales with the number density 
of the key--species, whereas the destruction rate ($J_i^{\rm dec}$) is proportional to the
vapour pressure of the molecule containing the key--element in equilibrium with the solid
particles.

\begin{table*}
\begin{center}
\caption{Dust species considered in the present analysis, their formation reaction and
the corresponding key elements (see text).} 
\begin{tabular}{l|l|c|c}
\hline
\hline 
Grain Species & Formation Reaction & Key element & Sticking coefficient\\
\hline
Olivine & 2$x$Mg +2(1-$x$)Fe+SiO+3H$_2$O $\rightarrow$ Mg$_{2x}$Fe$_{2(1-x)}$SiO$_4$ + 3H$_2$ & Si & 0.1 \\ 
Pyroxene & $x$Mg +(1-$x$)Fe+SiO+2H$_2$O  $\rightarrow$ Mg$_{x}$Fe$_{(1-x)}$SiO$_3$ + 2H$_2$ & Si & 0.1 \\
Quartz & SiO + H$_2$O $\rightarrow$ SiO$_2(s)$ +H$_2$ & Si & 0.1 \\  
Corundum & 2AlO+H$_2$O  $\rightarrow$ Al$_2$O$_3$ + H$_2$ & Al & 0.1-1 \\
Silicon Carbide & 2Si + C$_2$H$_2$ $\rightarrow$ 2 SiC + H$_2$ & Si & 1 \\ 
Carbon & C $\rightarrow$ C$(s)$ & C & 1 \\
Iron & Fe $\rightarrow$ Fe$(s)$ & Fe & 1 \\
\hline 
\hline 
\end{tabular}
\end{center}
\label{tabrates}
\end{table*}

The growth of the grain size of species $i$ is calculated via the equation

\begin{equation}
{da_i\over dt}=V_{0,i}(J_i^{\rm gr}-J_i^{\rm dec}),
\label{eqagrain}
\end{equation}  

\noindent
where $V_{0,i}$ is the volume of the nominal molecule in the solid.

The vapour pressures needed to estimate the decay rates $J_i^{\rm dec}$ were found 
by calculating the free enthalpy of formation of the molecules involved. Most of the
computations are based on the release by \citet{sharp90}, with the only exceptions of  
carbon and SiC, that were kindly provided by Prof. Gail (private communication).
Following \citet{fg06}, we do not consider any destruction reaction for solid carbon, but
we assume that it can form only at temperatures below 1100K.

The growth rates $J_i^{\rm gr}$ are proportional to the sticking coefficients $\alpha_i$
of the molecule including the key--elements on the solid grains. The choice of various
coefficients is the same as in our previuos investigations. For corundum, owing to lack
of laboratory data, we explore the range 0.1--1.

The composition of olivine and pyroxene depends on the relative fraction of magnesium
and iron, indicated, respectively, with $x$ and $1-x$ (see Table 1). The magnesium 
percentages in the olivine ($x_{ol}$) and pyroxene ($x_{py}$) dust particles are found
via the equations

\begin{equation}
{dx_{\rm ol}\over dt}={3V_{0,\rm ol}\over a_{ol}}\left[ 
(x_g-x_{\rm ol})J_{\rm ol}^{\rm gr}+{1\over 2}(J_+^{\rm ex}-J_-^{\rm ex}) \right]
\label{eqxol}
\end{equation}

\begin{equation}
{dx_{\rm py}\over dt}={3V_{0,\rm py}\over a_{py}}\left[ 
(x_g-x_{\rm py})J_{\rm py}^{\rm gr}+{1\over 2}(J_+^{\rm ex}-J_-^{\rm ex}) \right]
\label{eqxpy}
\end{equation}
\noindent
where $x_g$ is defined as the relative gas abundance of magnesium with respect to the Mg+Fe sum.

The quantity $(J_+^{\rm ex}-J_-^{\rm ex})$ gives the difference between the exchange
rate of iron by magnesium per unit surface area during collisions of Mg with the grain
surface and the rate of the reverse reaction, and is defined as in \citet{gs99}.

The amount of dust formed is indicated by the fraction of gas condensed into solid
grains. This is expressed by the fraction $f_i$ of the key--element that is in the solid
state. We have

\begin{equation}
f_i={4\pi (a_i^3-a_{0,i}^3)\over 3 V_{0,i}}{\epsilon_d\over \epsilon_i},
\end{equation}

\noindent
where $a_{0,i}$ is the initial size of the seed nuclei in the wind, $\epsilon_i$
is the number density of the key--elements in the wind, normalized to the hydrogen
density, and $\epsilon_d$ is the normalized density of the seed nuclei.

We assumed for each species $a_{0,i}=0.01 \mu$m; the final results are independent of
this assumption, provided that a miminum amount of dust, sufficient to accelerate the wind, 
is formed.

The situation for $\epsilon_d$ is more tricky. The typical number densities of grains in 
the outflow of M--stars of solar metallicity has been determined observationally, and is 
of the order of $10^{-13}$ \citep{knapp85}. The situation for C--stars and for M--stars of
different metallicity could be different (see the discussion in Nanni et al., 2013), but
presently any choice for $\epsilon_d$ would have a large degree of arbitrariness.

The opacity coefficient is found via the equation

\begin{equation}
k=k_{\rm gas}+\sum_{i}f_ik_i 
\end{equation}
where the $k_i$'s are the opacity coefficients for the various dust species formed.
The optical constants used to determine the opacities are the same as in \citet{paperI}.

The coupled system of differential equations for the wind and the dust formation of the
various species is solved with an Adams--Bashforth integrator \citep{golub92}. The
equations are solved in the radial outward direction. All the r.h.s. of eqs. \ref{eqagrain}
are set to zero until the first dust species become stable.

\label{refmodel}

\begin{figure*}
\begin{minipage}{0.3\textwidth}
\resizebox{1.\hsize}{!}{\includegraphics{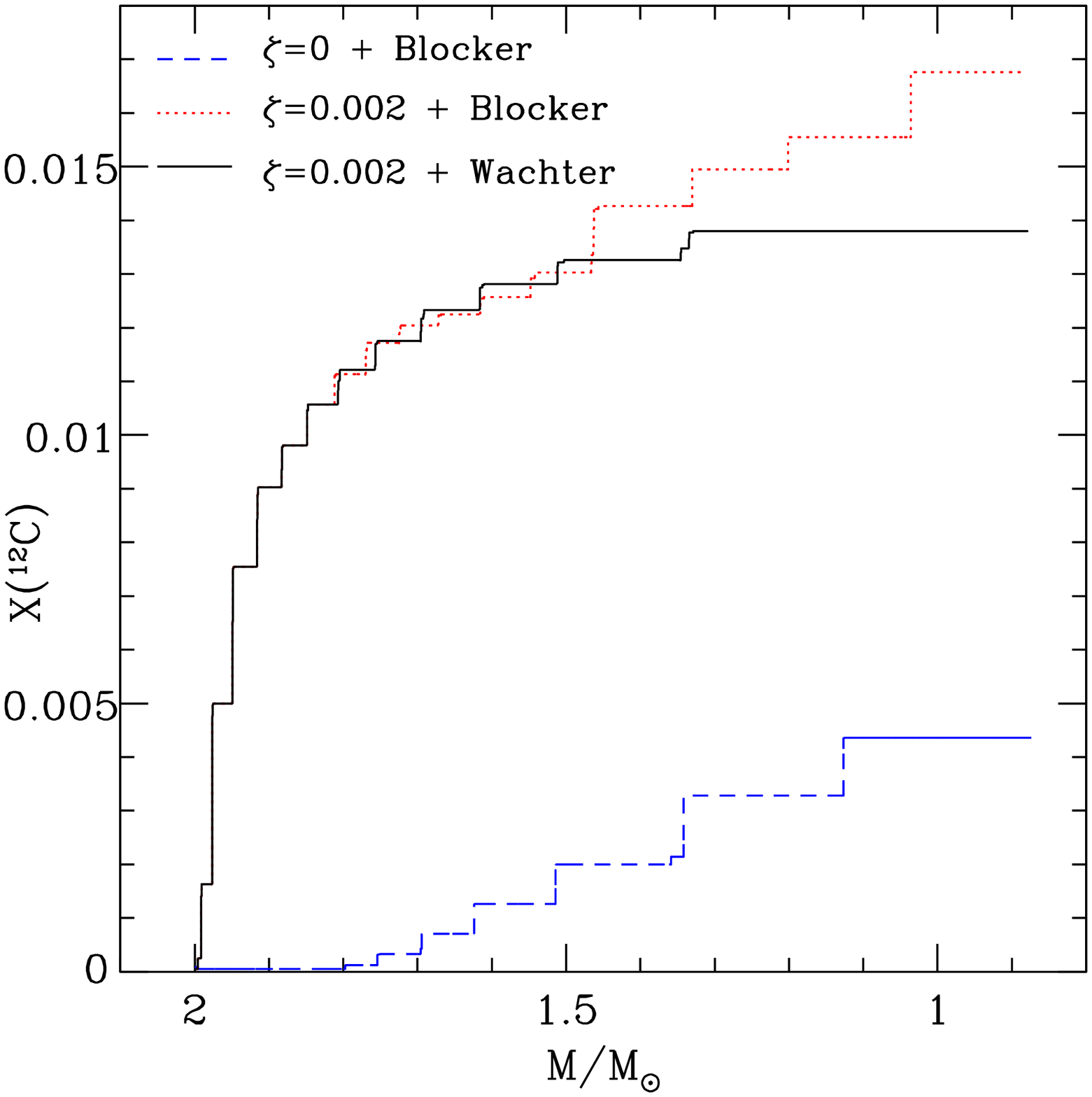}}
\end{minipage}
\begin{minipage}{0.3\textwidth}
\resizebox{1.\hsize}{!}{\includegraphics{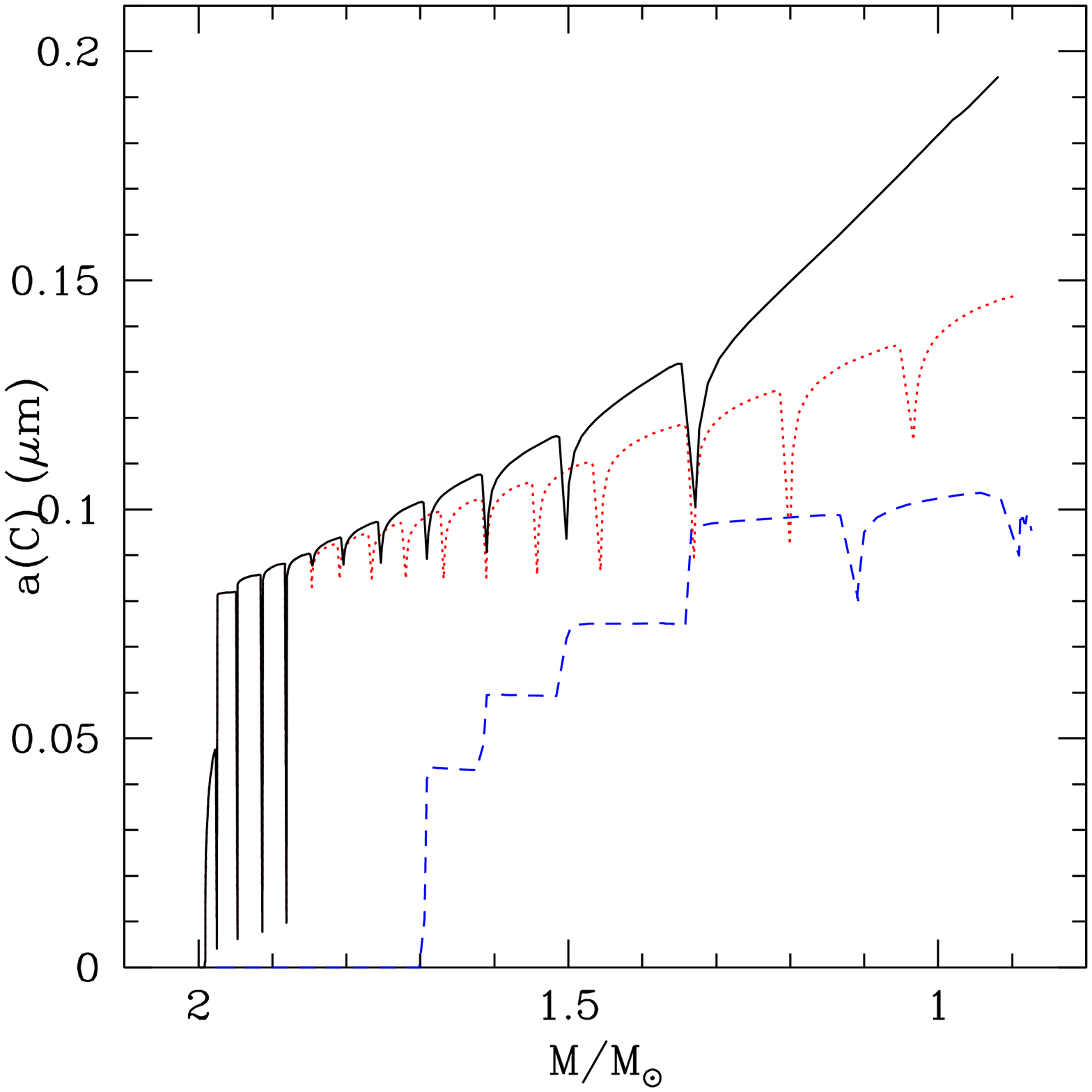}}
\end{minipage}
\begin{minipage}{0.3\textwidth}
\resizebox{1.\hsize}{!}{\includegraphics{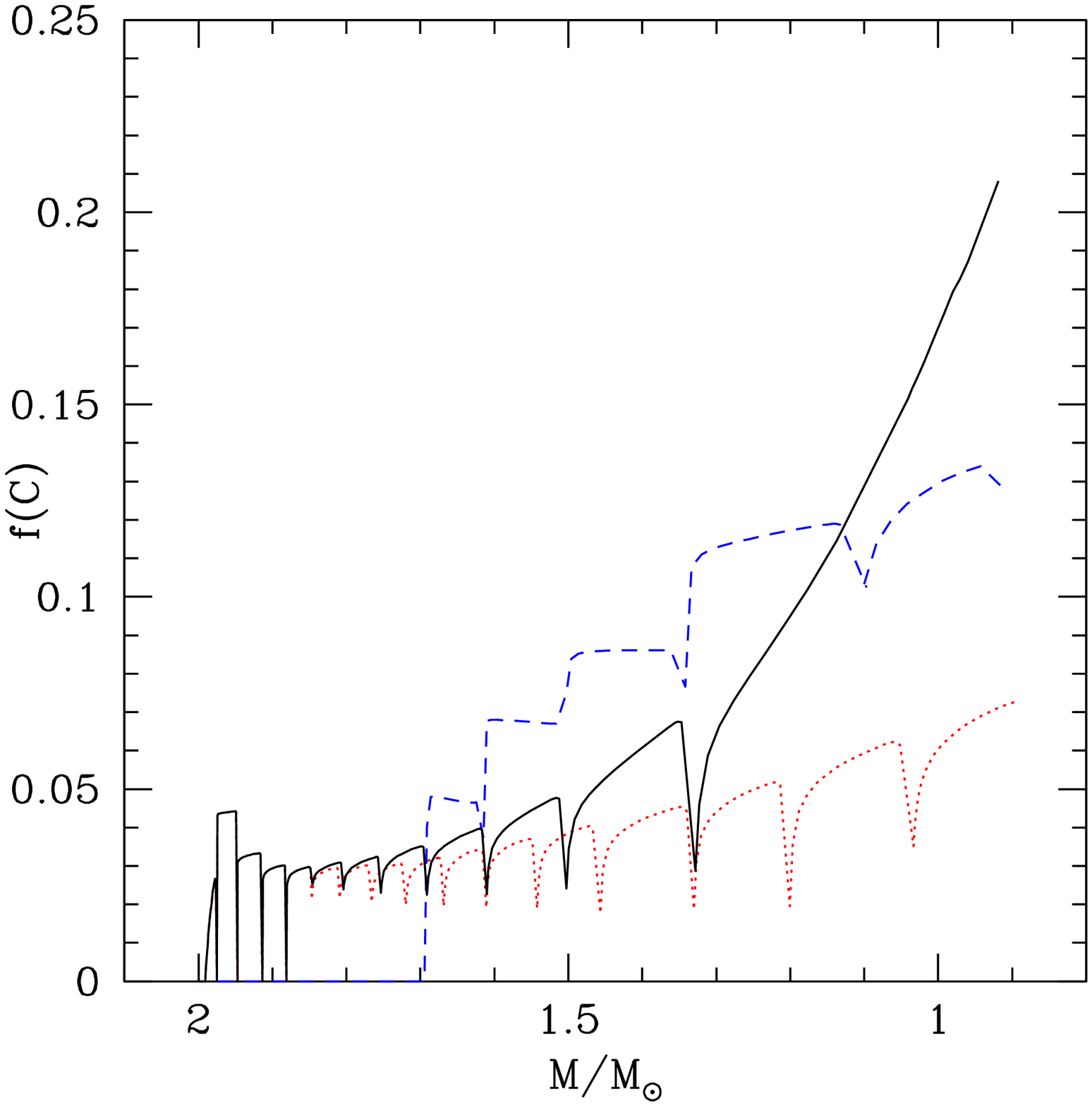}}
\end{minipage}
\begin{minipage}{0.3\textwidth}
\resizebox{1.\hsize}{!}{\includegraphics{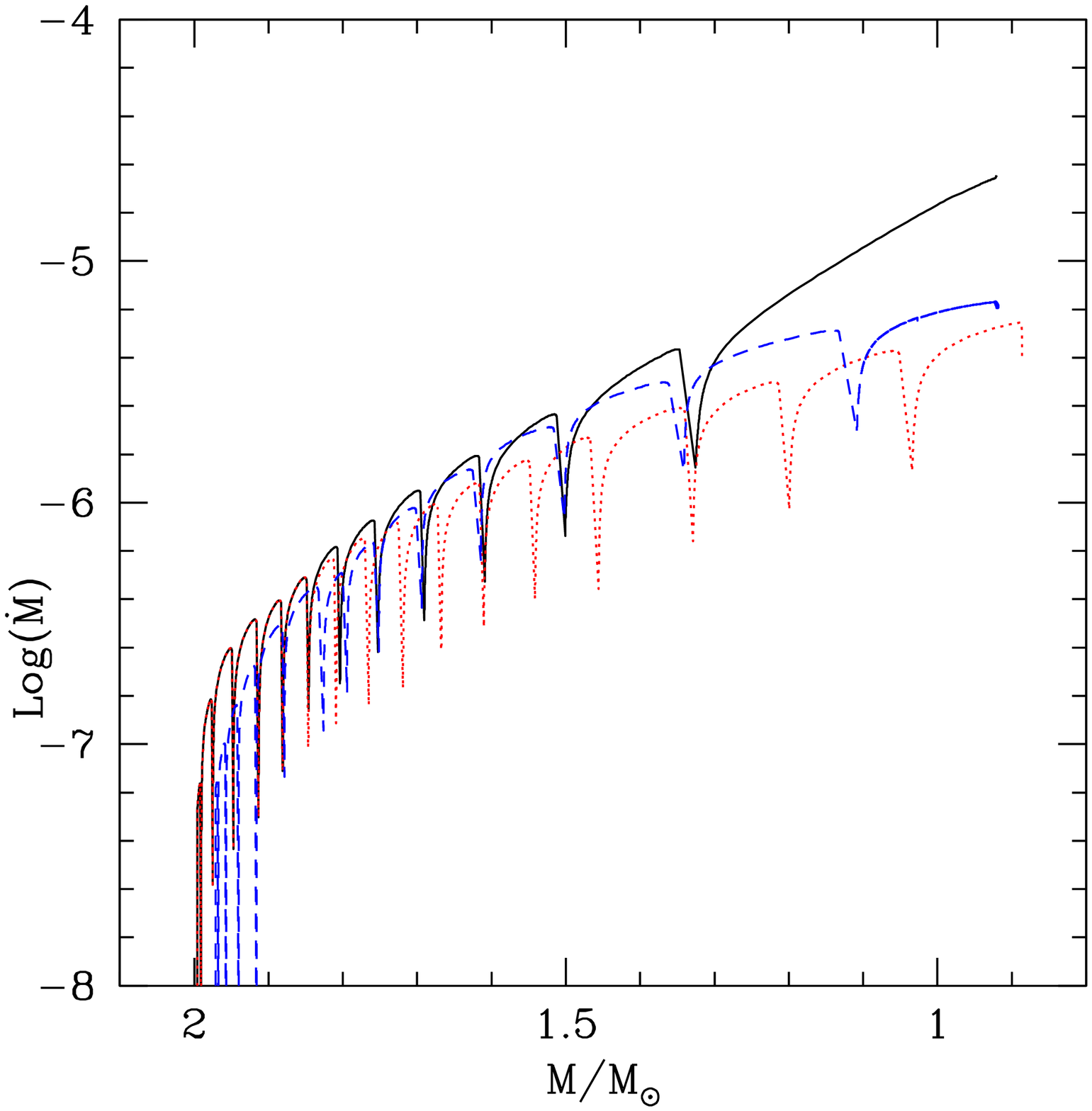}}
\end{minipage}
\begin{minipage}{0.3\textwidth}
\resizebox{1.\hsize}{!}{\includegraphics{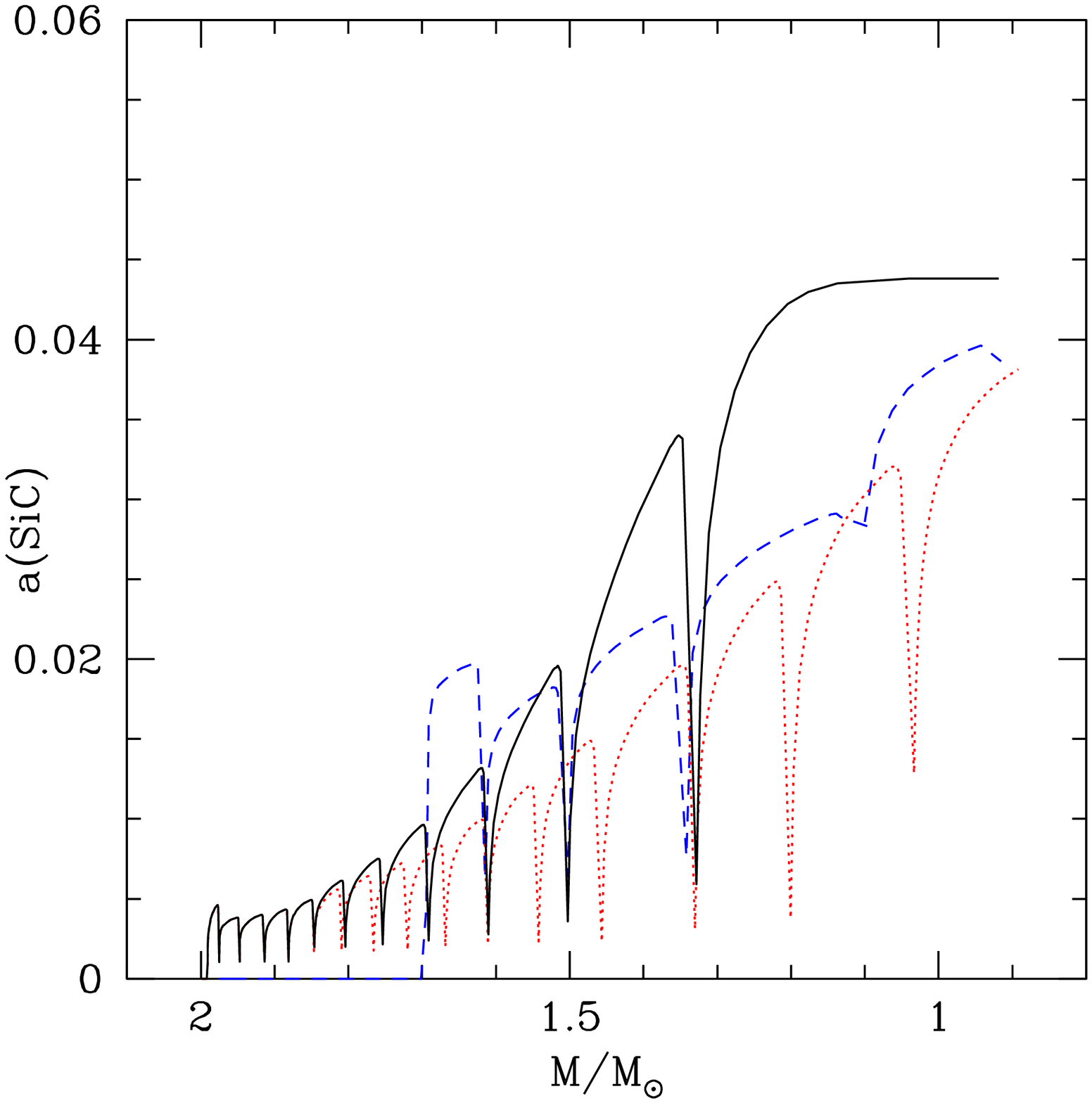}}
\end{minipage}
\begin{minipage}{0.3\textwidth}
\resizebox{1.\hsize}{!}{\includegraphics{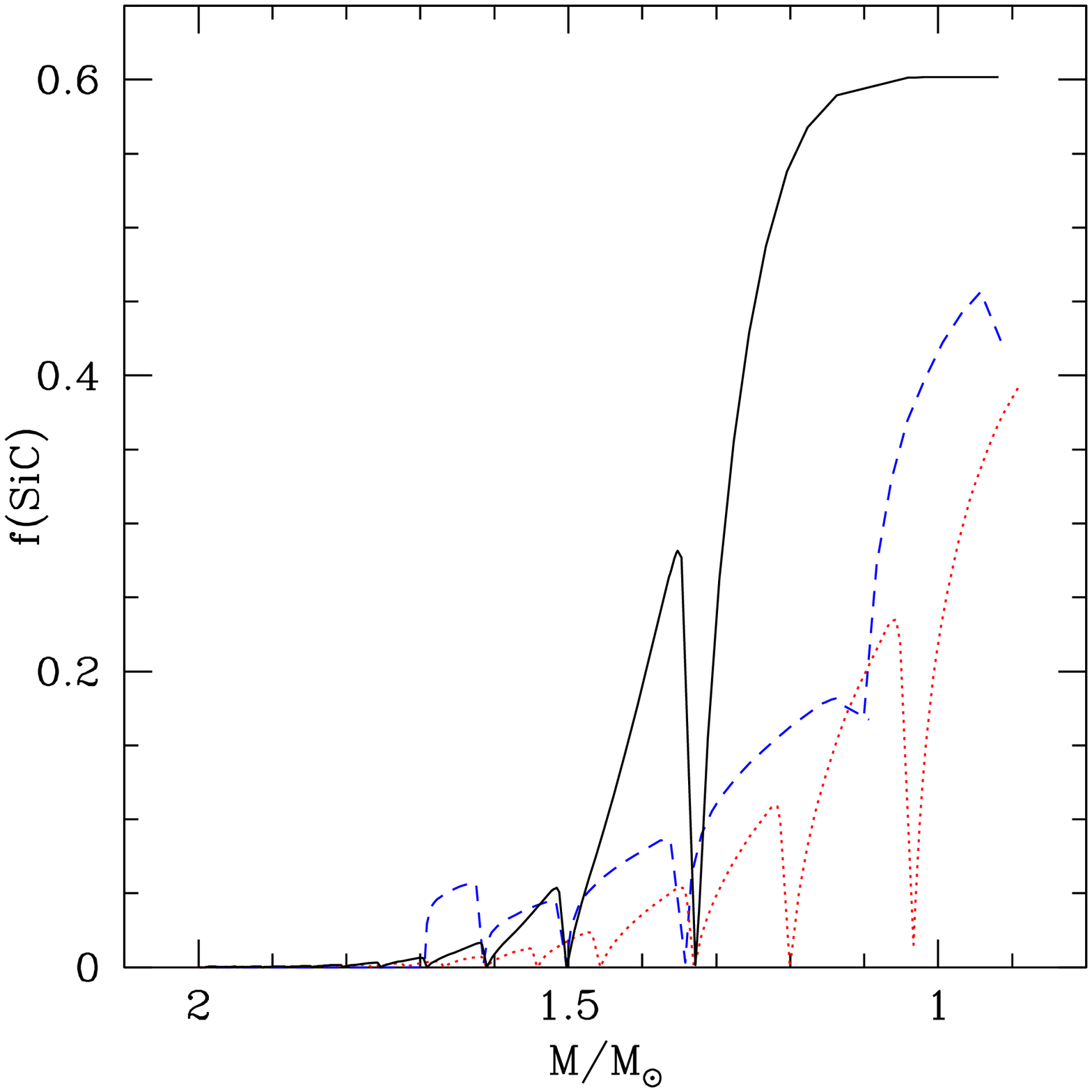}}
\end{minipage}
\vskip+10pt
\caption{The variation during the AGB evolution of a model of initial mass 2M$_{\odot}$,
of metallicity Z=0.001 \citep{paperI}. In the abscissa we indicate the current mass of the star,
decreasing during the evolution. In the top panels we show the carbon surface mass
fraction (left), the grains size of solid carbon dust particles (middle), and the
fraction of gaseous carbon condensed into dust (right). In the bottom panel
we report the mass loss rate (left), the size of SiC grains (middle), and the fraction
of silicon condensed into SiC dust. The models calculated with the Bl\"ocker mass loss
rate with no extra--mixing and with $\zeta=0.002$ are indicated, respectively, with blue,
dashed lines and with red, dotted tracks. The black, solid lines indicate models with the
mass loss description by \citet{wachter08}.
}
\label{2msun}
\end{figure*}

\section{Carbon dust production in AGBs}
\subsection{Current uncertainties from AGB modelling}
Models with M$\leq3$M$_{\odot}$ form silicates at the beginning of the AGB evolution, 
whereas in
more advanced phases repeated TDU episodes lead to C/O$>1$. The surface carbon enrichment
depends on the treatment of convective borders, particularly on the extension of the
extra--mixed region: this zone is stable, based on the Schwarzschild criterium,
yet it is reached by convective eddies, that cross the convective/radiative interface,
pushed by inertia.

\begin{figure*}
\begin{minipage}{1.0\textwidth}
\resizebox{1.\hsize}{!}{\includegraphics{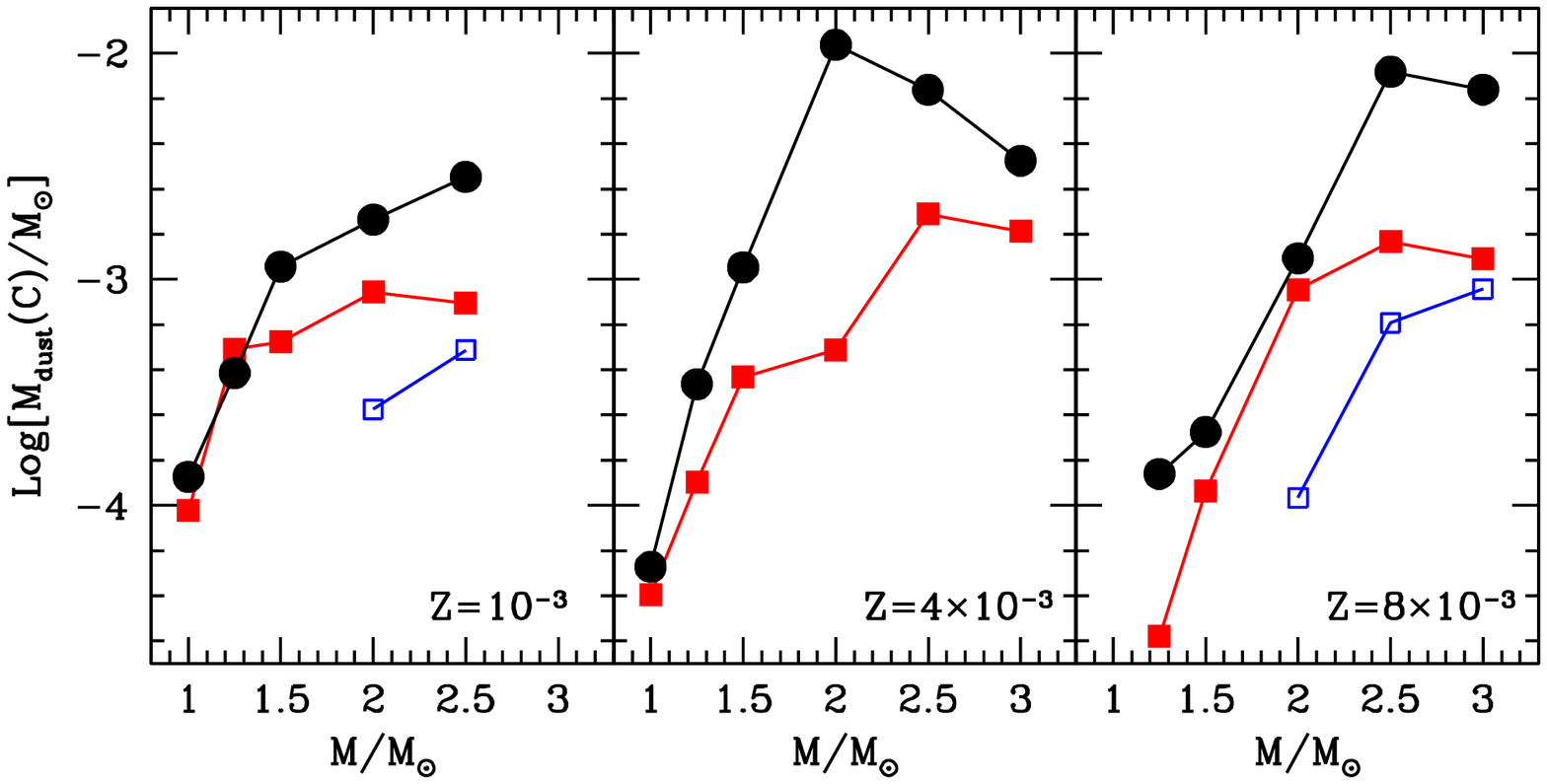}}
\end{minipage}
\vskip-230pt
\caption{Mass of solid carbon produced as a function of the initial mass of the star
for the metallicities $\rm{Z}=10^{-3}$ (left), $\rm{Z}=4\times 10^{-3}$ (middle), 
$\rm{Z}=8\times 10^{-3}$ (right). The meanings of the symbols is as follows.
Blue, open squares: no--overshoot models with the Bl\"ocker description of mass loss,
published in \citet{paperI, paperII}; 
Red, full squares: $\zeta=0.002$ models (Bl\"ocker's mass loss); 
Black, full circles: $\zeta=0.002$ models with the \citet{wachter08} mass loss.
}
\label{fcarb}
\end{figure*}

Locating the borders of convection zones during the AGB phase is one of the long--standing
problems in the context of stellar evolution \citep{mowlavi99}. Certain trends have
been identified by extant models, such as the increase in the extent of TDU in lower
metallicity models \citep{boothroyd88b}, and the fact that TDU ceases when the mass of the 
envelope drops below a threshold value \citep{straniero97}. Yet the extension of mixed
zones is still unknown from first principles: comparison with the observations of carbon stars
in the Galaxy \citep{wallerstein98} and in the Magellanic Clouds \citep{bessel83} indicate
that standard models, where mixing is allowed only within zones satisfying the Schwarzschild
criterium, fail to reproduce the observed patterns, and that some extra--mixing is required.

The presence of extra--mixing favours the increase in the surface carbon, for two reasons:
a) overshoot from the bottom of the convective envelope during the inwards penetration that
follows the thermal pulse makes more carbon--rich matter to be mixed with the surface 
layers; b) extra--mixing from the borders of the convective shell that forms during each TP
increases the strength of the pulse, in turn increasing the extent of the inwards
penetration of the convective envelope \citep{herwig04}.

Previous investigations fixed the conditions for TDU to occur either by imposing the
temperature of the innermost layer reached by the penetration of the convective mantle
\citep{marigo99}, or by assuming a relation between the core mass at which TDU begins as
a function of mass and metallicity \citep{karakas02, izzard04, marigo07}. In our case
the most straightforward way of introducing some overshoot is by assuming a non vanishing 
value of $\zeta$ (see Eq. 1).

The impact of assuming some extra--mixing on dust production during the AGB phase was
discussed on qualitative grounds in \citet{paperII}, by analysing the role played by 
$\zeta$ (see Eq. 1); no calibration of the extra--mixing was given. In this paper we 
use $\zeta=0.002$, according to the calibration of $\zeta$ aimed at reproducing the
luminosity function of carbon stars in the Magellanic Clouds, given in section 
3.1\footnote{It is not surprising that this value is a factor 10 smaller than
the $\zeta$ used to describe overshoot from convective cores during the H--burning phase,
given the different physical conditions in the two cases, and the much shorter duration 
of each individual TDU, compared to the time scales of core H--burning.}.

\begin{figure*}
\begin{minipage}{1.0\textwidth}
\resizebox{1.\hsize}{!}{\includegraphics{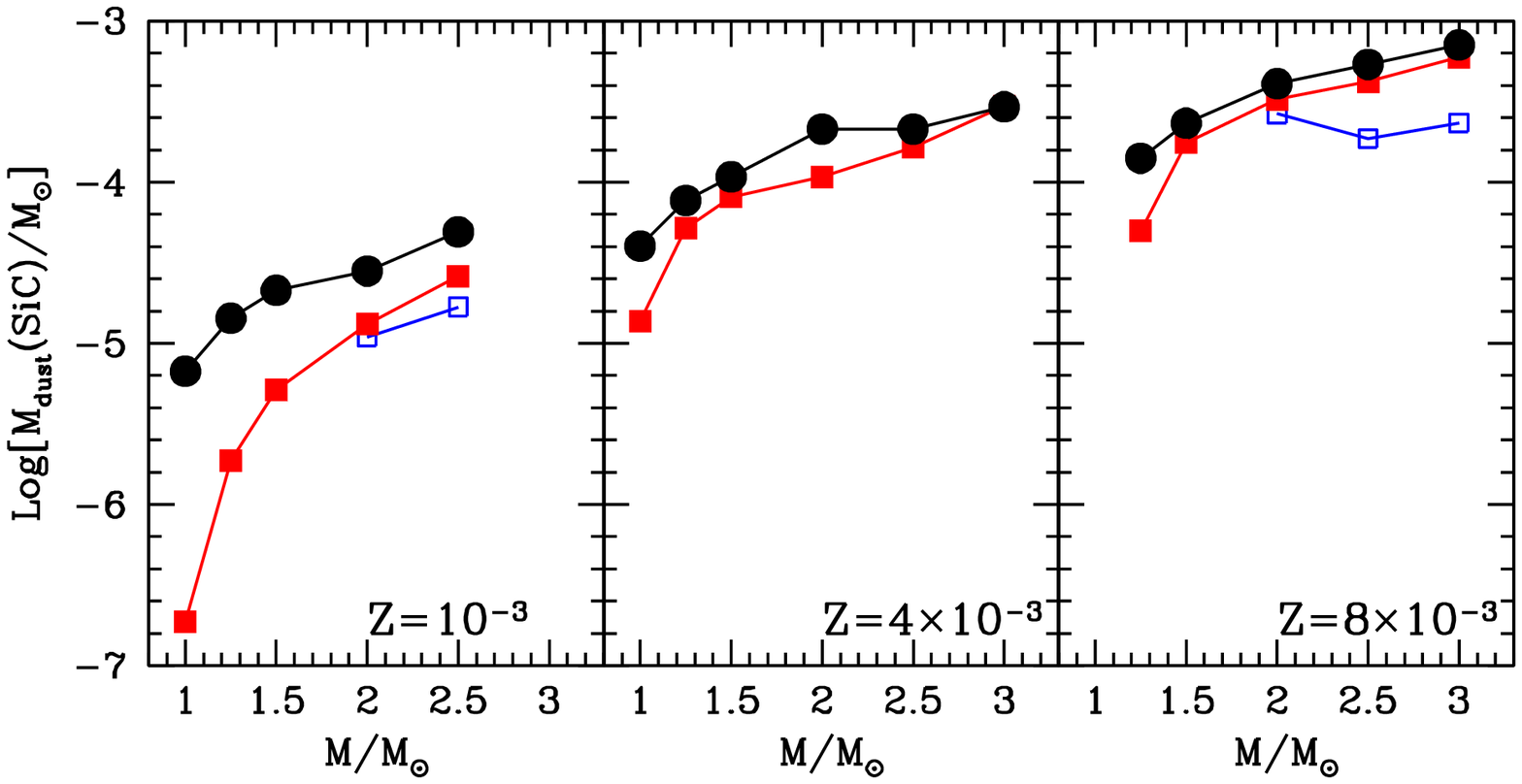}}
\end{minipage}
\vskip-230pt
\caption{Mass of silicon carbide produced by AGB models of different initial mass. 
The three panels refer to results for $\rm{Z}=10^{-3}$ (left), $\rm{Z}=4\times 10^{-3}$ 
(middle) and $\rm{Z}=8\times 10^{-3}$ (right). The meaning of the symbols is the same
as in Fig. \ref{fcarb}.
}
\label{fsic}
\end{figure*}

The effects of extra--mixing are shown in Fig. \ref{2msun}, where we compare the results 
obtained with no extra--mixing (blue, dashed tracks), with those found with $\zeta=0.002$
(red, dotted). We will refer to these models as {\it noov} and {\it over}, respectively.
The figure refers to the evolution of a 2M$_{\odot}$ model of metallicity Z=0.001,
published in \citet{paperI}.

In the upper, left panel, we see that the carbon increase is much more pronounced in the
{\it over} model : the surface carbon reaches a final mass fraction of X(C)$\sim 0.017$,
a factor $\sim 3$ larger than in the {\it noov} case. This has a straight consequence on
the production of solid carbon, as we see in the upper, middle panel, showing the
variation of the solid carbon size in the two cases: the size of carbon grains formed
in the {\it over} model approaches $a_C \sim 0.15 \mu $m, whereas in the {\it noov}
case they barely reach $a_C \sim 0.1 \mu $m. Carbon dust formation occurs at
an earlier phase in the {\it over} model, because TDU is more efficient in favouring the
transition from M-- to C--star; in the {\it noov} case, the star looses 0.2M$_{\odot}$
before the condition C/O$> 1$ is reached.

\begin{figure*}
\begin{minipage}{1.0\textwidth}
\resizebox{1.\hsize}{!}{\includegraphics{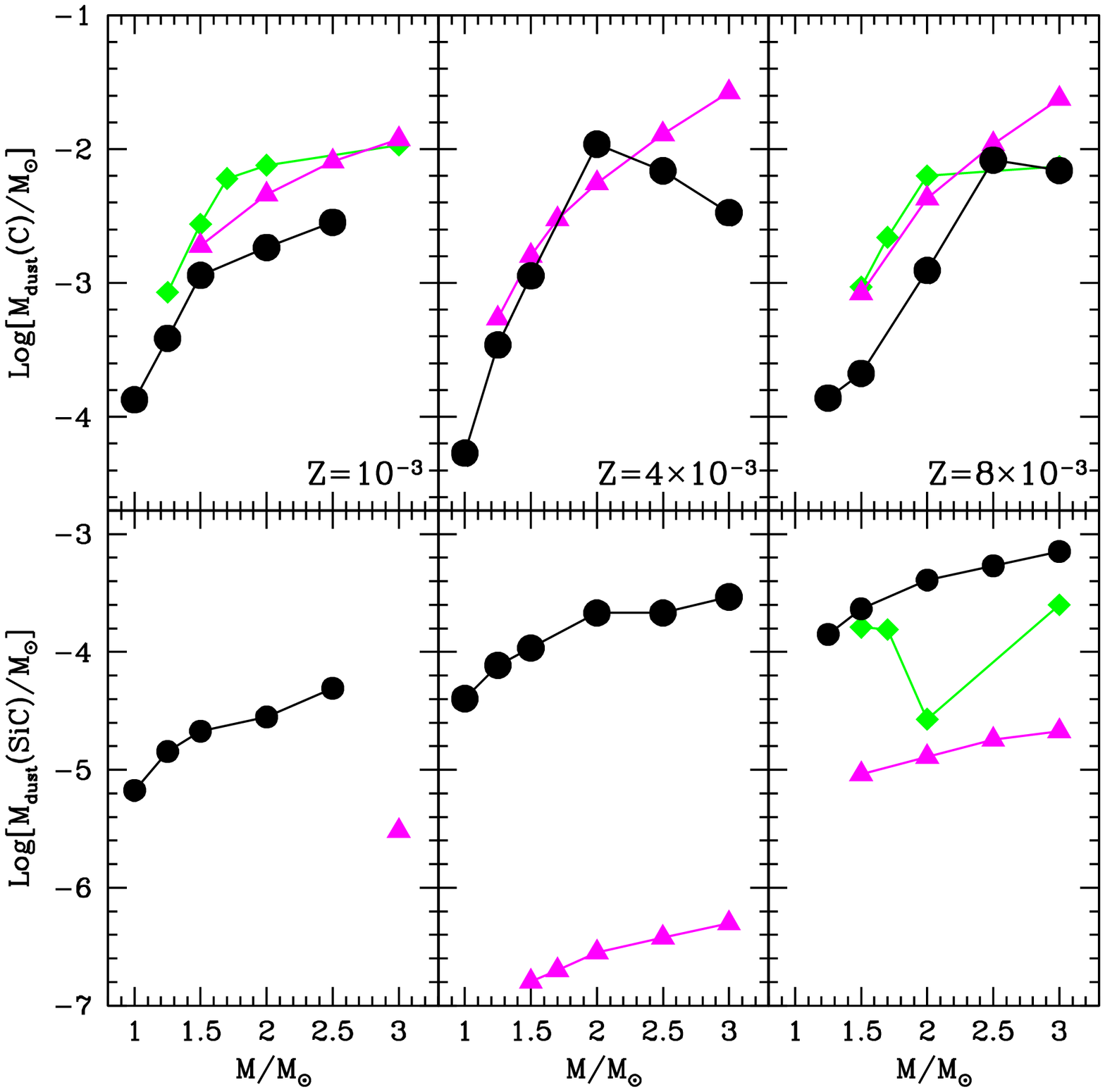}}
\end{minipage}
\vskip-5pt
\caption{The comparison among the mass of solid carbon (top panels) and SiC (bottom)
produced by AGB models of various initial mass and metallicities Z$=10^{-3}$ (left),
Z$=4\times 10^{-3}$ (middle), Z$=8\times 10^{-3}$ (right). The meaning of the symbols is
as follows: Black circles: models with mass loss by \citet{wachter08} and $\zeta=0.002$; 
Magenta triangles: models by \citet{fg06}; Green diamonds: models by \citet{nanni13}.
}
\label{fconfrontiC}
\end{figure*}

The fraction of carbon condensed into dust (right, upper panel) is larger in the
{\it noov} case, owing to the smaller amount of carbon available at the surface.

The quantity of silicon carbide formed is less sensitive to the treatment of convective
borders, as shown in the middle and right, lower panels of Fig. \ref{2msun}. This is 
because the key--element for SiC is silicon, whose surface content is 
independent of the efficiency of TDU. Clearly production of SiC demands that the
C--star stage has already been achieved, thus SiC grains start to form in earlier phases
in the {\it over} model.

To discuss the role played by the description of mass loss,
we compare the results obtained with the
treatment based on the Bl\"ocker's formula with the more recent compilation by 
\citet{wachter08}. This latter is 
more suitable to describe the mass loss mechanism in these evolutionary phases, as
it is based on simulations of stellar winds driven by radiation pressure on carbon grains.
The rates by \citet{wachter08} show a great sensitivity to the effective temperature of
the star, and show a rapid increase in the rate of mass loss as soon as 
carbon--dust formation begins.

The results obtained with the \citet{wachter08} formula are indicated in Fig. \ref{2msun} 
with black, solid, tracks. We used the same prescription of the 
extra--mixing from convective borders as for the {\it over} case, thus we can analyse
the role of mass loss treatment by comparing
with the red, dotted lines in the six panels of Fig. \ref{2msun}. 

The mass loss rate given by the \citet{wachter08} formula are larger once the C--star
stage is reached, as can be seen in the left, bottom panel: the star experiences a smaller 
number of TPs, thus the surface carbon enrichment (shown in the left, upper panel) is lower.
Despite the lower carbon available, more dust is formed: this is due to the larger density 
of the wind, favoured by the higher mass loss rate.

We see in the middle, bottom panel that the higher $\dot M$ also favours a greater 
production of SiC.

The overall comparison among results obtained with the different assumptions concerning 
the convective borders and the mass loss mechanism are shown in Fig. \ref{fcarb} (solid
carbon produced) and in Fig. \ref{fsic} (silicon carbide). The three panels refer to
the metallicities Z=$10^{-3}$ (left), Z=$4\times 10^{-3}$ (middle), Z=$8\times 10^{-3}$
(right). Part of the results concerning Z=$10^{-3}$ and Z=$8\times 10^{-3}$ were
published in \citet{paperI, paperII}. Here we extend the computations to models
calculated with the mass loss by \citet{wachter08}, and to the metallicity 
Z=$4\times 10^{-3}$; for this latter Z the case $\zeta=0$ with no extra--mixing was not 
explored.

In the Z=$10^{-3}$ case we see that when extra--mixing is neglected only models of
initial mass M=$2,2.5$M$_{\odot}$ become carbon stars and produce solid carbon. Use of the 
mass loss rate by \citet{wachter08}
favours a larger dust production for M$\geq 1.5$M$_{\odot}$. For smaller masses
the effect of the higher mass loss rates is counterbalanced by the lower carbon
available, such that the results turn out to be approximately independent of the
mass loss description. 

The Z=$4\times 10^{-3}$ models show a similar behaviour, the impact of mass loss
increasing for larger masses. The amount of solid carbon produced (and also
the differences determined by the mass loss treatment) reaches a maximum around for stars 
of initial mass around $2,2.5$M$_{\odot}$, and becomes progressively smaller as the
mass increases, and approaches the threshold limit for HBB ignition.

\begin{figure*}
\begin{minipage}{1.0\textwidth}
\resizebox{1.\hsize}{!}{\includegraphics{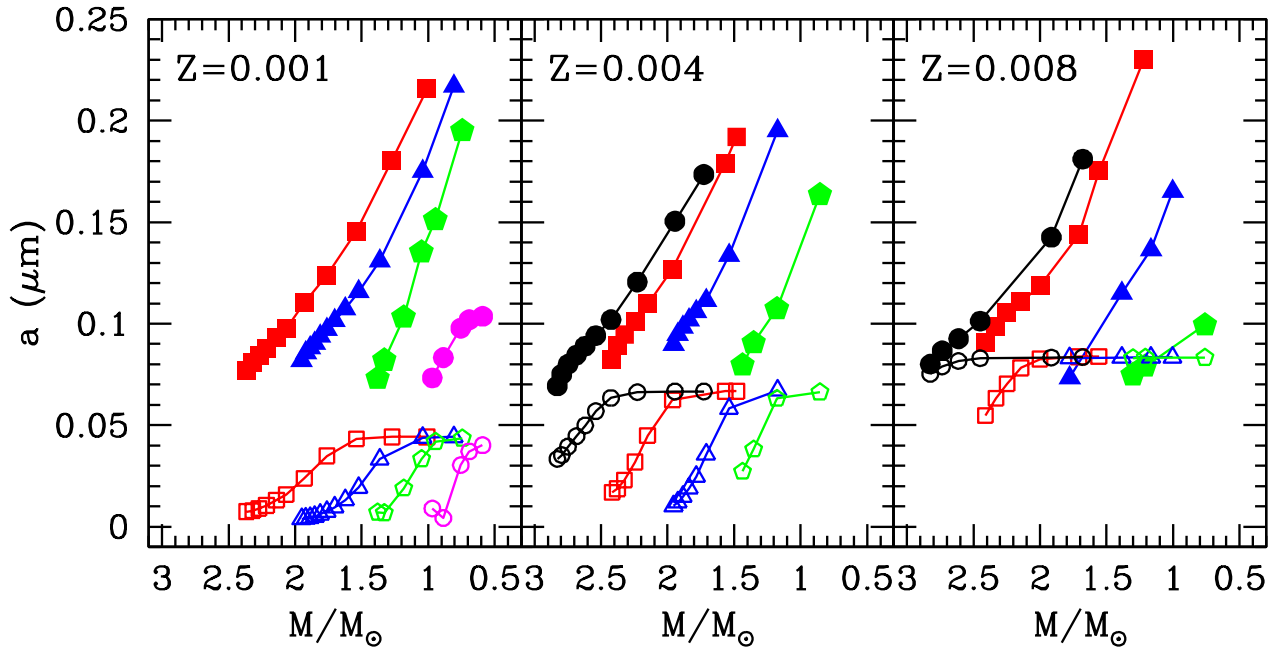}}
\end{minipage}
\vskip-230pt
\caption{The variation of the size of dust grains of carbon (full points) and
SiC (open) during the AGB evolution of models of metallicity ${\rm Z}=10^{-3}$ (left panel),
${\rm Z}=4\times 10^{-3}$ (middle), ${\rm Z}=8\times 10^{-3}$ (right). The initial masses
are $3$M$_{\odot}$ (black circles), $2.5$M$_{\odot}$ (red squares), $2$M$_{\odot}$ (blue 
triangles), $1.5$M$_{\odot}$ (green pentagons). For the ${\rm Z}=10^{-3}$ metallicity
we also show the $1$M$_{\odot}$ model (magenta circles).
}
\label{fgrains}
\end{figure*}

At Z=$8\times 10^{-3}$, similarly to Z=$10^{-3}$, we find that when no 
overshoot is considered only models with M$\geq 2$M$_{\odot}$ produce solid carbon. 
Use of the mass loss prescription by \citet{wachter08}
favours an increase in the total solid carbon produced of almost one order of magnitude
for masses above $\sim 2$M$_{\odot}$. For smaller masses the differences are negligible,
because use of the \citet{wachter08} formula favours a rapid loss of the envelope 
as soon as the C--star is achieved, which prevents further carbon to be dredged--up to 
the surface.

Fig. \ref{fsic} shows that the production of silicon carbide is also dependent on the
treatment of convective borders and the description of mass loss. The results are 
easier to be interpreted in this case, owing to the approximately constant value of the
key--element, i.e. silicon, which, unlike carbon, is not changed by TDU.

Models using the mass loss treatment by \citet{wachter08} produce more silicon carbide,
owing to the higher density in the wind. Clearly models with no overshoot produce a
smaller amount of SiC, because they reach the C--star stage in later phases, after
part of the envelope was lost.

\subsection{The comparison with other investigations}
We compare the results by \citet{fg06}, and the more recent investigation by \citet{nanni13} 
with our models. This comparison is aimed to stress the differences in the AGB
modelling, because the three studies use the same description of the thermodynamics
of the wind and of the dust formation process.

In Fig. \ref{fconfrontiC} we indicate the results by \citet{fg06} and
\citet{nanni13}, respectively, with magenta, full triangles and green, solid diamonds.
The comparison for Z=$4\times 10^{-3}$ is limited to \citet{fg06}, as this metallicity
was not discussed by \citet{nanni13}.

In all cases the amount of carbon dust formed increases with mass, because more massive
stars experience more thermal pulses, dredge--up more carbon at the surface, and also
loose the external envelope at a higher rate, which increases the number density of gas
particles in the wind.

For Z=$10^{-3}$ the carbon dust
produced by our models for M$\geq 2$M$_{\odot}$ is a factor of $\sim 3$ smaller
compared to \citet{fg06}. The interpretation of such a difference is not straightforward,
because the models by \citet{fg06} are based on a synthetic modelling, and keep the same
efficiency of TDU until the latest AGB phases, when the mass of the envelope is
greatly reduced. Also, in our models we account for the modification of the surface
chemistry in the computation of the low--temperature opacities via the AESOPUS tool
\citep{marigo09}: this leads to a general cooling of the outer layers, which, in turn,
favours the expansion of the external regions, a higher mass loss rate, and a faster
consumption of the stellar envelope, before a great enrichment of surface carbon can be
achieved \citep{vm09, vm10}. The smaller densities of carbon nuclei is probably the 
main reason for the lower amount of solid carbon produced by our models \citep{paperI}. 

The comparison with the results by \citet{nanni13} also shows that
the carbon dust produced by our models is lower, although the differences 
are smaller than with \citet{fg06}. These can be ascribed to 
the choice of the initial density of seed nuclei: we assumed 
$\epsilon_d=10^{-13}$, whereas \citet{nanni13} scale $\epsilon_d$ with the carbon
excess with respect to oxygen \citep[see session 4.1 in][]{nanni13}.

Turning to the metallicity of the new models presented here, Z$=4\times 10^{-3}$,
we see in the top, middle panels of Fig. \ref{fconfrontiC} that the difference with \citet{fg06}
is smaller than for Z=$10^{-3}$. The models by \citet{fg06} produce more dust
for M$\geq$2.5M$_{\odot}$; this is due to the different convective model adopted,
because the use of the FST description partially limits the efficiency of TDU for
the masses close to the limit for HBB.

For $Z=8\times 10^{-3}$ the differences among the three compilations depends on the
stellar mass. For M$\leq 2$M$_{\odot}$ our yields are a factor $\sim 3$ smaller
compared to \citet{fg06} and \citet{nanni13}. These models
experience a strong mass loss as soon as they become carbon stars, owing to the
large sensitivity of the formulation by \citet{wachter08} to the effective temperature 
of the star, and the small temperatures typical of AGBs of this metallicity. Consequently, 
they loose their 
envelope very rapidly, preventing the possibility that large quantities of carbon are
dredged--up to the surface. For masses above $\sim 2$M$_{\odot}$ the carbon dust produced
by our models is similar to \citet{nanni13}, and smaller than \citet{fg06}; this 
likely reflects the difference in the computation of the low--T opacities in the
C--rich mixture, as discussed previously.

The amount of silicon carbide formed is independent of the surface carbon enrichment,
because in this case silicon is the key--element. We see in the bottom panels of 
Fig. \ref{fconfrontiC} that, independently of metallicity, more SiC is formed in our case, 
compared to \citet{fg06} and \citet{nanni13}. The reason for this is in the details of the
growth of SiC and solid carbon grains in the winds of C--stars. SiC is formed in more
internal regions, and is rather transparent to the radiation, thus determining only a 
modest acceleration of the wind. Carbon grains are formed in more external layers, 
and their large opacity favours a fast acceleration of the wind. The formation of
carbon grains halts the growth of SiC grains, because the acceleration of the wind
determines a drop in the density of silicon particles in the wind. In the models by
\citet{fg06} and \citet{nanni13} the amount of carbon formed is larger, which, in turn,
determines a lower content of SiC.

\subsection{Dust from C--stars: the trend with mass and metallicity}
In the previous sections we stressed how dust production
by C--stars is sensitive to many uncertainties, the most relevant being the
treatment of convective borders and the description of mass loss.

We consider as our reference model the results obtained with the extra--mixing during the
AGB phase calibrated as discussed in section 3.1, and the mass loss treatment 
modelled as in \citet{wachter08}. These results are indicated as black, full circles in
Fig. \ref{fcarb}, \ref{fsic}, \ref{fconfrontiC}.

Fig. \ref{fgrains} shows the variation during the AGB life of the grain size of dust
particles formed around stars of various masses and metallicities. Full points 
indicate the dimension of solid carbon grains, whereas open points refer to SiC.
The choice of the current mass of the star as abscissa helps to understand 
how many grains of a given size are formed around the star during the whole evolution. 
Each point refers to
the grain size in the middle of each interpulse phase. For the metallicities
$Z=4\times 10^{-3}$ and $Z=8\times 10^{-3}$ we show results for models of initial masses
1.5, 2, 2.5, 3M$_{\odot}$; in the $Z=10^{-3}$ case, since the 3M$_{\odot}$ model shows
that HBB is active and that the C--star stage is not reached, we show the 1M$_{\odot}$ 
case instead.

Concerning solid carbon, we see that for each mass the size of dust particles,
a$_C$, increases during the evolution, owing to the increase in the surface carbon abundance 
and in the mass loss rate. Values of $a_{\mathrm C}$ span the range 0.07--0.2 $\mu$m, 
although for masses M$\leq 1.5M_{\odot}$ we find $a_{\mathrm C}\leq 0.1\mu$m.

We see in the three panels of Fig. \ref{fcarb} that in the low--mass regime, with 
M$<2$M$_{\odot}$, the amount of carbon dust formed increases with mass, independently 
of $Z$: the higher is M, the larger is the surface carbon enrichment. In this interval
of mass the $Z=10^{-3}$ models produce more carbon, because low--$Z$ models experience 
deeper TDUs \citep{boothroyd88a}, and reach more easily the C--star stage, owing to the 
smaller content of oxygen initially present at the surface.

For masses M$\geq 2$M$_{\odot}$ dust production is favoured in the higher metallicity
models, because these evolve at smaller effective temperatures: dust formation occurs in 
layers closer to the stellar surface, where the densities are higher.

The mass of solid carbon produced, m$_{\mathrm C}$, ranges from $10^{-4}$M$_{\odot}$
to $\sim 5\times 10^{-3}$M$_{\odot}$ at $Z=10^{-3}$, whereas at higher metallicities
m$_C \sim 10^{-2}$M$_{\odot}$ for the two masses M=$2,2.5$M$_{\odot}$.

In the three bottom panels of Fig. \ref{fgrains} we see that SiC grains grow during the 
evolution until a threshold size is reached, after which no further growth 
occurs. The maximum dimension achieved by SiC grains ranges from $\sim 0.05\mu$m
for $Z=10^{-3}$, to $\sim 0.07\mu$m for $Z=4\times 10^{-3}$, up to
$\sim 0.09\mu$m for $Z=8\times 10^{-3}$.

This behaviour can be explained by the saturation due to the lack of silicon particles
in the wind. Owing to the stability of SiS molecule, in carbon--rich environments
we have n$_{\rm Si}=[(1-f_{\rm SiC})\epsilon_{\rm Si}-\epsilon_{\rm S}]$n$_{\rm H}$
\citep{fg06}, where $\epsilon_{\rm Si}$ and $\epsilon_{\rm S}$ are the number density of 
silicon and sulphur particles in the wind (normalized to the hydrogen density), whereas 
f$_{\rm SiC}$ is the fraction of silicon condensed into SiC. This expression poses an
upper limit on the amount of SiC that can form, corresponding to the value of f$_{\rm SiC}$
at which n$_{\rm Si}$ vanishes:

$$
{\rm f}_{\rm SiC}=1-{\epsilon_{\rm S} \over \epsilon_{\rm Si}}
$$

\noindent
Because in solar or $\alpha-$enhanced mixtures 
${\epsilon_{\rm S} \over \epsilon_{\rm Si}} \sim 0.4$ \citep{grevesse98}, f$_{\rm SiC}$ cannot
exceed $\sim 60\%$.

The mass of SiC (m$_{\rm SiC}$) produced increases with the stellar mass, as can be seen
in Fig. \ref{fsic}, because of the larger mass returned into the interstellar medium 
by massive AGBs during the evolution.
Higher--metallicity models produce more SiC, owing to the larger amount
of silicon present in the stars. $Z=10^{-3}$ models produce an amount of SiC
$10^{-5}$M$_{\odot} < $m$_{\rm SiC}<10^{-4}$M$_{\odot}$, whereas at $Z=4\times 10^{-3}$
we find $5\times 10^{-5}$M$_{\odot} < $m$_{\rm SiC}<5\times 10^{-4}$M$_{\odot}$;
the largest production of SiC occurs at $Z=8\times 10^{-3}$, for which we have
$10^{-4}$M$_{\odot} < $m$_{\rm SiC}<10^{-3}$M$_{\odot}$. The scaling of the mass of
SiC produced with the metallicity is approximately linear, as expected.

\section{The effects of Hot Bottom Burning on massive models}

\begin{figure}
\begin{minipage}{0.45\textwidth}
\resizebox{1.\hsize}{!}{\includegraphics{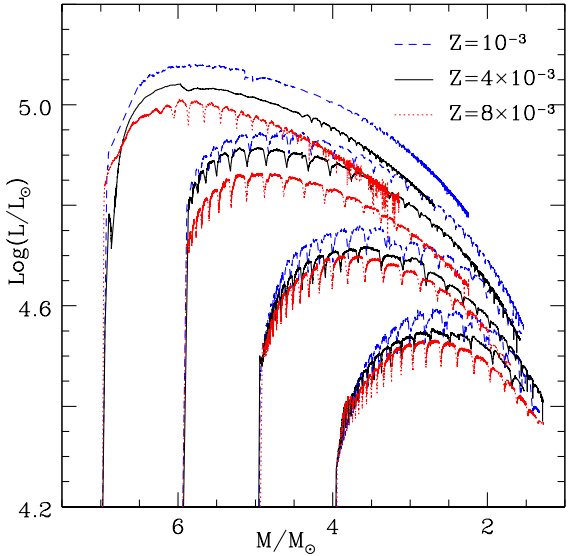}}
\end{minipage}
\vskip-00pt
\caption{The AGB evolution of the luminosity of models of initial
mass M$\geq 4$M$_{\odot}$ for the metallicities $Z=10^{-3}$ (blue, dashed lines),
$Z=4\times 10^{-3}$ (black, solid) and $Z=8\times 10^{-3}$ (red, dotted track).
}
\label{flumhbb}
\end{figure}

Stellar models with mass above $\sim 3$M$_{\odot}$ experience HBB during the interpulse
phases of the AGB evolution. The key--parameter to describe the HBB strength is the
temperature at the bottom of the convective envelope, which determines the
degree of nucleosynthesis experienced in those regions, and the modification
of the surface chemistry. Stars that experience HBB reach very high luminosities \citep{blocker91}, 
much larger than predicted by the classic relation by \citet{paczynski}.

Fig. \ref{flumhbb} shows the evolution of the luminosity of AGB models of various 
initial masses and metallicities. 

The luminosity first increases in the early AGB
phases, then declines until the envelope is consumed. The initial trend is due to
the increase in the core mass of the star, whereas the consumption of the envelope is
the reason for the later decrease in the overall energy flux.

Higher mass models evolve on bigger cores, experience stronger HBB, and evolve
at larger luminosities.

Models with smaller metallicity are substained by a more powerful CNO--burning shell:
HBB conditions are reached more easily in lower $Z$ models, that evolve at larger
luminosities. We see in Fig. \ref{flumhbb} that for a given mass the 
$Z=4\times 10^{-3}$ models (hereinafter {\rm z4m3}, see Ventura et al. (2013b) for a more
exhaustive discussion of the physical properties of AGB models with $Z=4\times 10^{-3}$) 
presented here evolve at a luminosity intermediate between models of the same mass of 
metallicity $Z=10^{-3}$ (hereinafter {\rm z1m3}) and $Z=8\times 10^{-3}$ ({\rm z8m3}).

\begin{figure*}
\begin{minipage}{1.0\textwidth}
\resizebox{1.\hsize}{!}{\includegraphics{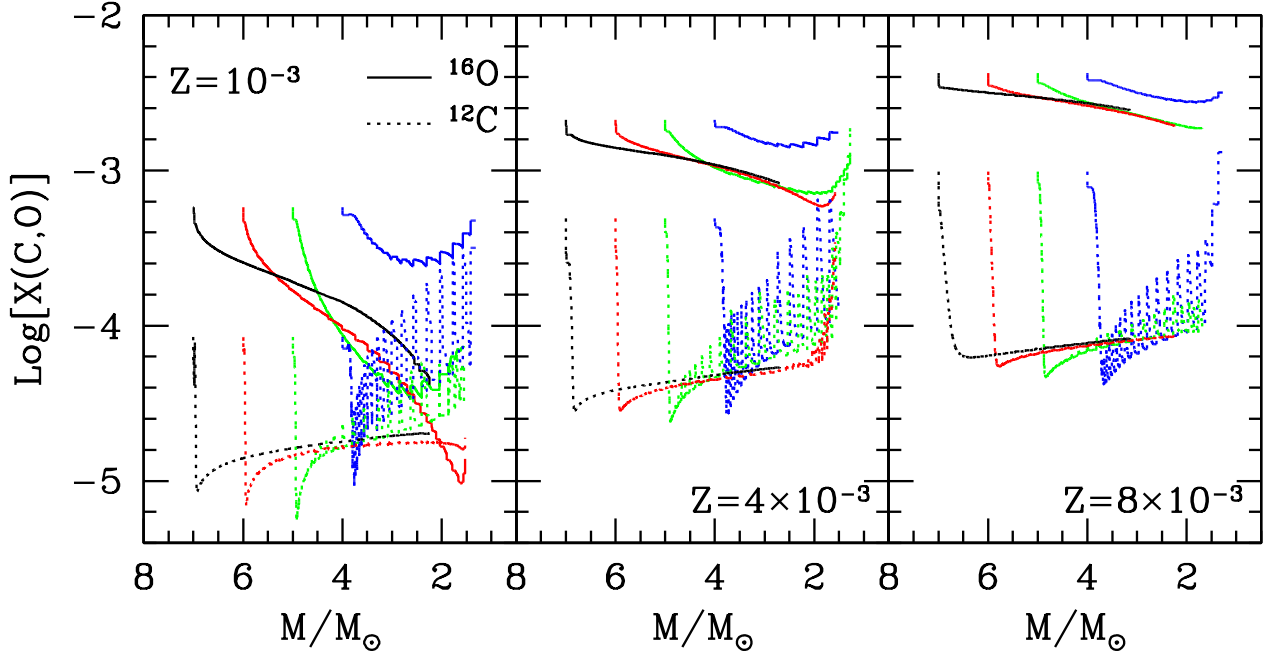}}
\end{minipage}
\vskip-230pt
\caption{The evolution of the the surface mass fraction of carbon (dotted lines) and
oxygen (solid tracks) in models of initial mass in the range $4-8$M$_{\odot}$ of
metallicity $Z=10^{-3}$ (left), $Z=4\times 10^{-3}$ (middle) and 
$Z=8\times 10^{-3}$ (right). The different colors correspond to masses
$7$M$_{\odot}$ (black), $6$M$_{\odot}$ (red), $5$M$_{\odot}$ (green),
$4$M$_{\odot}$ (blue).
}
\label{fcohbb}
\end{figure*}

The evolution of the surface chemistry of massive AGBs can also be explained on the
basis of the HBB experienced. Fig. \ref{fcohbb} shows the variation of the
surface abundances of oxygen (solid tracks) and carbon (dotted lines) for the same
models shown in Fig. \ref{flumhbb}. 

The strong depletion of the surface carbon in the early AGB phases is the signature of
HBB, starting at $\sim 40$MK. The depletion of oxygen requires higher temperatures at 
the bottom of the convective zone (T$_{\rm bce} \sim 70$MK), thus it takes place
in more advanced phases. The {\rm z8m3} models experience a soft HBB, thus the reduction 
of the surface oxygen is modest (at most a factor $\sim 2$). 
In the {\rm z1m3} models the depletion of the surface 
oxygen is not monotonic with mass: stars with mass around $\sim 6$M$_{\odot}$ produce 
the most O--poor ejecta, the initial oxygen being destroyed by almost two orders of
magnitude. More massive stars, though experiencing a 
stronger HBB, loose their surface envelope very rapidly, before a very advanced 
nucleosynthesis can be activated: their surface oxygen is slightly higher 
\citep{vd11, ventura13a}.

In all the cases shown in Fig. \ref{fcohbb} we see that the surface C/O keeps below
unity, thus preventing the possibility that the C--star stage is reached, and,
consequently, that carbon--type dust is formed.

\begin{figure*}
\begin{minipage}{1.0\textwidth}
\resizebox{1.\hsize}{!}{\includegraphics{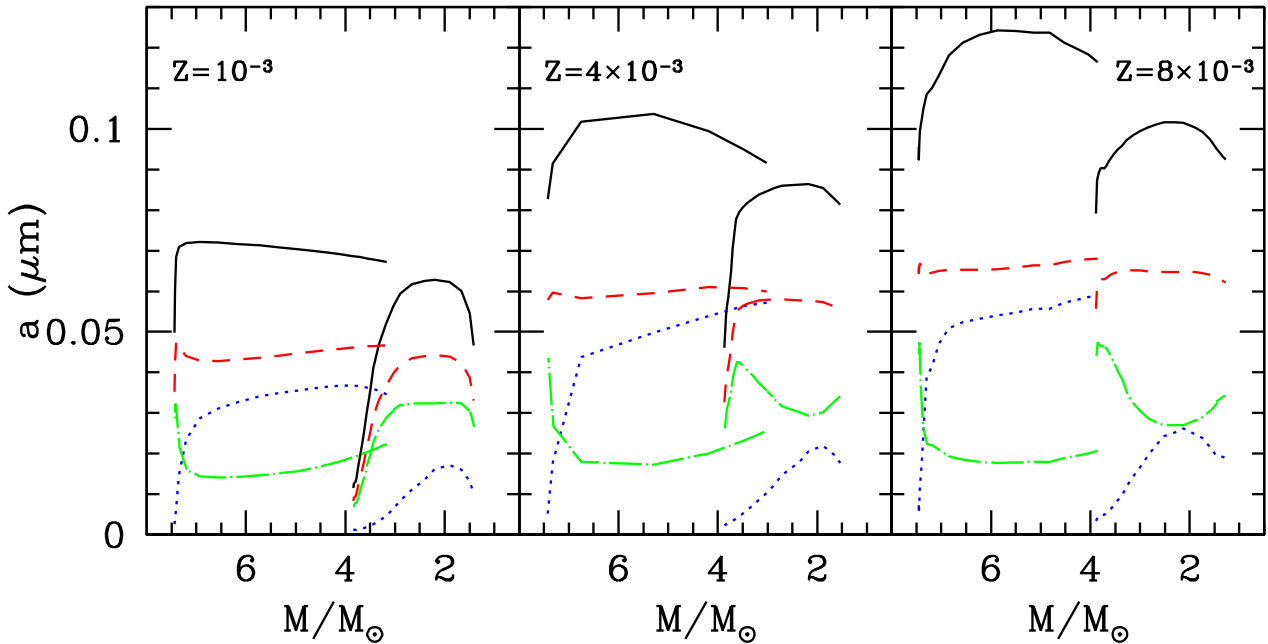}}
\end{minipage}
\vskip-230pt
\caption{The variation of the grains size of olivine (black, solid lines), pyroxene
(red, dashed), quartz (green, dot--dashed) and corundum (blue, dotted) in models of
initial mass $4$M$_{\odot}$ and $7.5$M$_{\odot}$ of metallicity $Z=10^{-3}$ (left), 
$Z=4\times 10^{-3}$ (middle) and $Z=8\times 10^{-3}$ (right). 
}
\label{fsizehbb}
\end{figure*}

The size of the dust grains of the various species that form during the AGB evolution
is shown in the three panels of Fig. \ref{fsizehbb}. For clarity reasons we show only the
$4$M$_{\odot}$ and $7.5$M$_{\odot}$ cases, as representative of the lowest and highest
masses experiencing HBB.

In agreement with previous investigations,
we find that olivine is the dominant species \citep{fg06}. For the {\rm z4m3} and {\rm z8m3}
cases the dimension of the olivine grains formed increases with the stellar mass. The 
grain size $a_{\rm ol}$ is in the range $0.07 \mu{\rm m} < a_{\rm ol} < 0.13 \mu$m for
$Z=8\times 10^{-3}$, and $0.06 \mu {\rm m} < a_{\rm ol} < 0.11 \mu$m for $Z=4\times 10^{-3}$.
The {\rm z4m3} produce smaller amounts of olivine than their {\rm z8m3} counterparts,
because although they evolve at larger luminosities and loose mass at a higher rate,
their surface silicon (which scales with the metallicity, and is scarcely touched by HBB)
is smaller. The {\rm z1m3} models produce even less olivine, $a_{\rm ol}$ 
never exceeding $\sim 0.07 \mu$m. The formation of
olivine grains in the more massive models of this metallicity stops at a certain stage 
during the AGB evolution, because the strong depletion of the surface oxygen (see the left 
panel of Fig. \ref{fcohbb}) prevents the formation of water molecules, that are essential 
to form olivine grains. This behaviour was found and discussed in \citet{paperI}. 

\begin{figure*}
\begin{minipage}{0.45\textwidth}
\resizebox{1.\hsize}{!}{\includegraphics{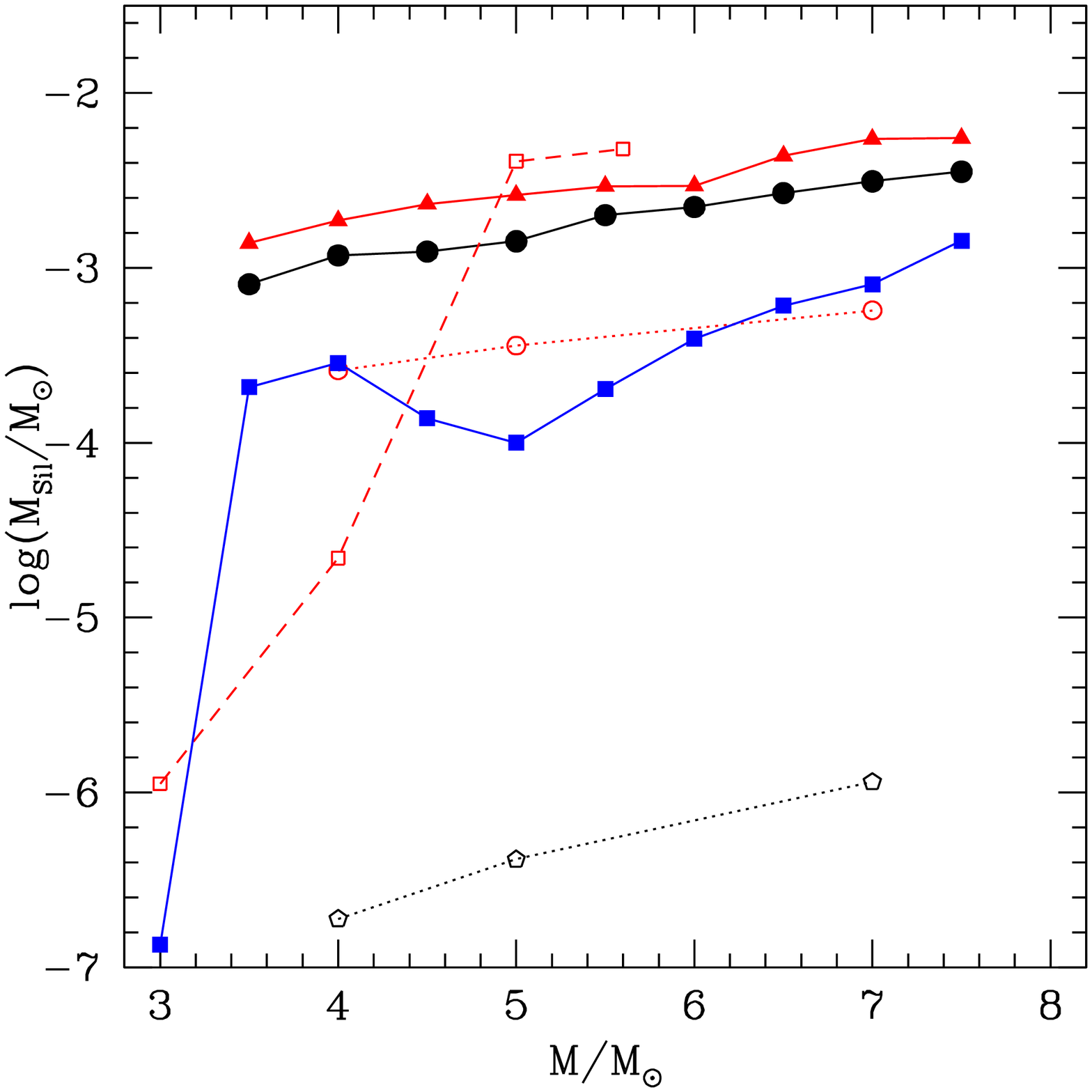}}
\end{minipage}
\begin{minipage}{0.45\textwidth}
\resizebox{1.\hsize}{!}{\includegraphics{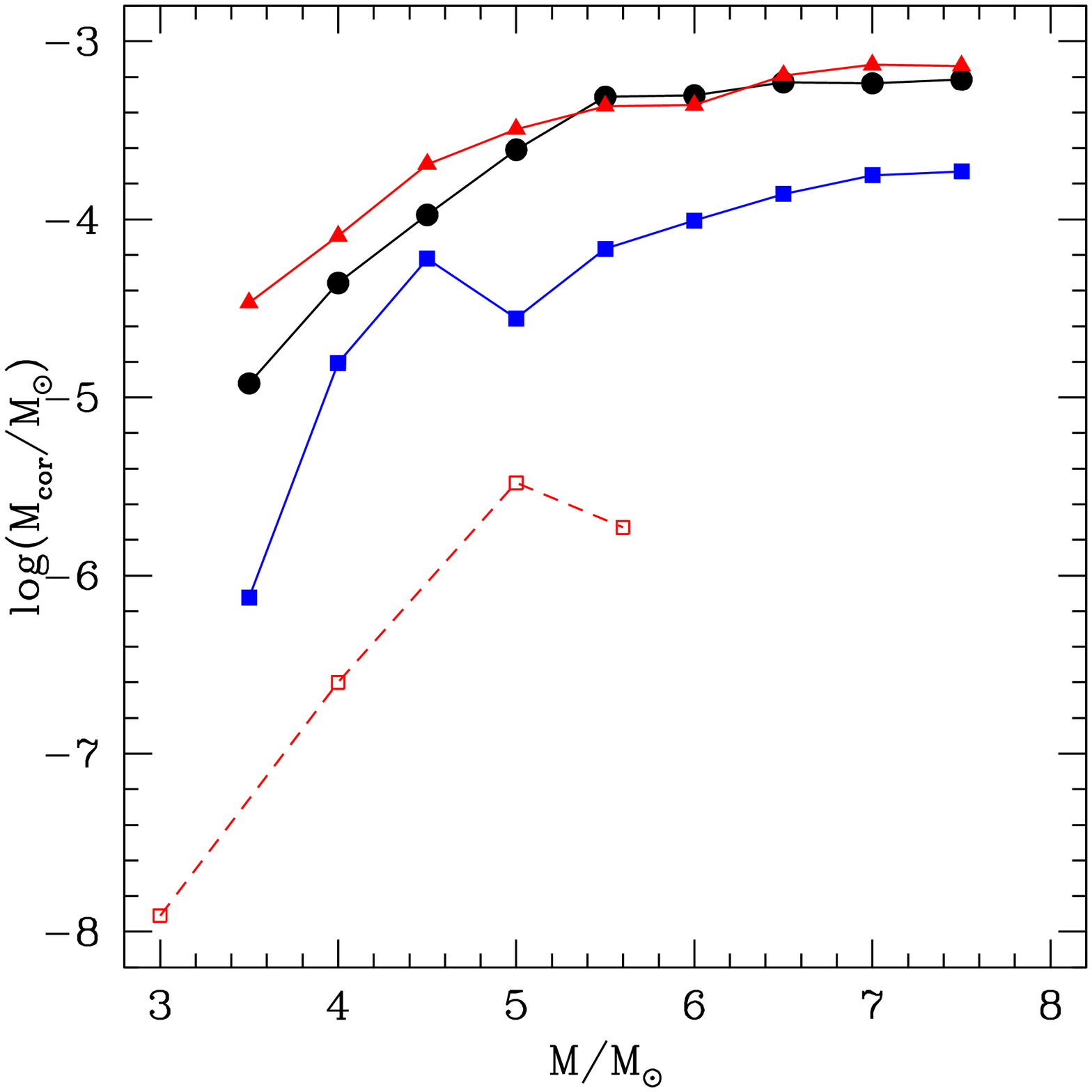}}
\end{minipage}
\vskip+10pt
\caption{Left: The mass of silicates produced as a function of the initial mass
for the metallicities $Z=8\times 10^{-3}$ (red, solid triangles),
$Z=4\times 10^{-3}$ (black, full circles), $Z=10^{-3}$ (blue, 
full squares). Open points refer to results from \citet{fg06} of metallicity
$Z=8\times 10^{-3}$ (red, open circles) and $Z=4\times 10^{-3}$ 
(black, open circles); these results are connected with dotted lines.
Open squares (connected with a dashed line) refer to results from \citet{nanni13}, 
of metallicity $Z=8\times 10^{-3}$. Right: The mass of corundum produced
by massive AGBs. The meaning of the symbols is the same as in the left panel.
}
\label{fsilhbb}
\end{figure*}

In the winds of M--stars pyroxene is the second most abundant species after olivine.
Depending on the metallicity, the size of pyroxene grains ranges from $\sim 0.04 \mu$m to 
$\sim 0.07 \mu$m (see Fig. \ref{fsizehbb}).

Quartz is the least abundant among the silicates, the grain size keeping below
$\sim 0.04 \mu$m. Note that in the early AGB phases the size of quartz grains is
larger: this is because a smaller amount of olivine is formed, thus the acceleration
experienced by the wind is smaller, which contributes to form more quartz.

Besides silicates, an additional species that forms in the winds of M--stars is corundum,
an Al--rich compound very stable \citep{sharp90}, and extremely transparent to 
radiation \citep{koike95}.
In our previous investigations the formation of corundum was neglected, although, owing to
its stability, it is likely that considerable amounts of this dust species from in
regions close to the surface of the star \citep{gs98}. The details of the corundum condensation
process and a discussion of the related uncertainties, will be published in a separate
paper (Dell'Agli et al., submitted). 

The variation of the corundum grain size during the
AGB evolution is different compared to silicates: corundum particles form in larger and larger
dimensions as the evolution proceeds, whereas the size of silicates follows the path
traced by luminosity (see Figg. \ref{flumhbb} and \ref{fsizehbb}). This is an effect of
HBB, which, via activation of the Mg--Al nucleosynthesis, favours a gradual aluminium 
enrichment of the surface layers \citep{ventura13a}.  

The total mass of silicates and corundum is shown in 
Fig. \ref{fsilhbb}. For the metallicities $Z=4\times 10^{-3}$ and $Z=8\times 10^{-3}$
the mass of silicates produced is in the range 
$10^{-3}$M$_{\odot} < $M$_{\rm Sil} < 10^{-2}$M$_{\odot}$ and increases with the stellar 
mass, in agreement with the previous discussion on the grain sizes of olivine and pyroxene.
The {\rm z1m3} models produce a smaller quantity of dust, owing to the scarcity of silicon available, 
but also because of the strong depletion of the surface oxygen: this is the reason for
the dip around $\sim 5$M$_{\odot}$, that can be seen in the {\rm z1m3} line in the left
panel of Fig. \ref{fsilhbb}.

The comparison with the results by \citet{fg06} shows that the mass of silicates produced 
by our models is much larger, owing to the stronger HBB experienced, which, in 
turn, favours a large increase in the mass rate, hence in the density of the wind. This is 
particularly evident at the small metallicities, where the production of dust by the
\citet{fg06} is modest, as a consequence of the scarcity of silicon in the surface layers
of the star (the models by \citet{fg06} of metallicity $Z=10^{-3}$ are not shown in the
figure, as they would fall off the chosen scale for the vertical axis).

The comparison with the results by \citet{nanni13} is less straightforward. At 
$Z=8\times 10^{-3}$ the overall mass of silicates produced by our models is larger, though
the two most massive models in the \citet{nanni13} compilation produce a quantity of
silicates comparable to ours. In the $Z=10^{-3}$ case no silicates is produced by
\citet{nanni13} models, because the models, though experiencing a soft HBB, eventually
become carbon stars, thus produce carbon--type dust.

A word of caution is needed here. In the models presented here we follow the same 
schematization described in \citet{fg06}, in that we assume that the disintegration of
olivine, and more generally of silicates, takes place by chemisputtering: \citet{gs99}
discuss the stability of silicates, arguing against the possibility that pure
thermal composition is responsible for the disintegration of silicates.  This
assumption is compatible with the silicates forming reactions given in Tab. \ref{tabrates}.
Conversely, \citet{nanni13} assume that the destruction of silicates grains is
triggered by a pure vaporization process. The difference between these two descriptions
is that while in the chemisputtering case silicates are not stable at temperatures
above $\sim 1100$K, in the second hypothesis this threshold is lifted to $\sim 1500$K.
Clearly in the latter case more silicates are formed, because the condensation process
begins in regions closer to the surface of the star, at larger densities.
Discriminating among these two hypothesis is beyond the scope of the present investigation,
however we are more favourable to the chemisputtering solution because: a) condensation
temperatures for silicates as high as 1500K are not observed in radiative transfer models
for O--rich circumstellar dust shells \citep{groenewegen09}; b) observationally, the highest
temperatures observed are $\sim 1350$K for aluminum dust \citep{karovicova13}.

The mass of corundum produced, $\rm m_{co}$, is also increasing with the stellar mass
(see right panel of Fig. \ref{fsilhbb}). Because of the saturation conditions reached, 
there is practically no difference between the $Z=4\times 10^{-3}$ models and their 
$Z=8\times 10^{-3}$ counterparts: $\rm m_{co}$ varies from a few $10^{-6}$M$_{\odot}$
for M$=3.5$M$_{\odot}$, to $\sim 10^{-3}$M$_{\odot}$ for M$=7.5$M$_{\odot}$. 
The $Z=10^{-3}$ models produce a smaller amount of corundum, owing to the low surface mass 
fraction of aluminium.

\begin{figure}
\begin{minipage}{0.45\textwidth}
\resizebox{1.\hsize}{!}{\includegraphics{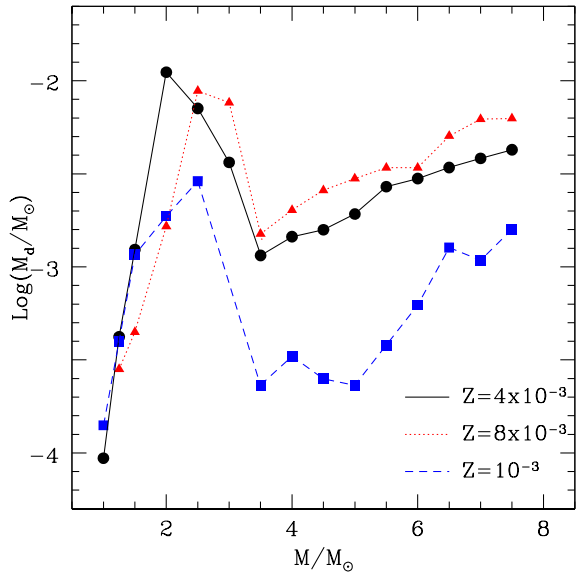}}
\end{minipage}
\vskip-00pt
\caption{The mass of dust produced by AGB models of various initial mass. The
various symbols refer to the metallicities $Z=10^{-3}$ (blue squares),
$Z=4\times 10^{-3}$ (black circles) and $Z=8 \times 10^{-3}$ (red triangles).
}
\label{fdusttot}
\end{figure}

\section{Conclusions}
In the previous sections we discussed separately the dust yields by stars with
mass ${\rm M}<3M_{\odot}$, that produce carbon dust, and by more massive AGBs,
that produce silicates. TDU favours the surface carbon enrichment
for the lower masses, where HBB dominates for masses above $3{\rm}M_{\odot}$.

The total dust produced by stars of various masses and metallicities is shown in
Fig. \ref{fdusttot}. For consistency with the new grid of AGB models with $Z=4\times 10^{-3}$
presented here, the mass of carbon dust produced by $Z=10^{-3}$ and $Z=8\times 10^{-3}$
(discussed in Ventura et al. 2012a,b) have been recalculated to account for: 
i) a small extra--mixing from the bottom of the convective envelope and ii) the mass loss rate by 
\citet{wachter08}, that is more suitable to describe mass loss for carbon stars.
Both these effects increase the quantity of carbon--dust formed around 
M$<3$M$_{\odot}$ AGBs. The extra--mixing below the base of the envelope favours a more 
penetrating TDU, thus a higher surface carbon available. Use of the \citet{wachter08}
mass loss leads to higher rates for carbon stars; this, in turn, determines an increase 
in the density of the wind, and consequently a higher rate of dust formation.

The quantity of dust formed depends on the initial mass of the star. The various lines
in Fig. \ref{fdusttot}, indicating different metallicities, outline an initial 
increasing trend for masses M$<2-2.5$M$_{\odot}$, owing to the higher number of TPs
experienced, which favour a larger carbon enrichment of the surface layers. In these
stars $\sim 80-90\%$ of dust is under the form of solid carbon, with $\sim 10-20\%$
of SiC and solid iron.

The stars with mass around $3.5$M$_{\odot}$ produce smaller quantities of dust,
because HBB prevents the formation of carbon--type dust, and their mass loss
rates are not sufficiently large to trigger the formation of large amount of silicates.

Higher mass models produce more dust, because they evolve at larger luminosities,
thus they loose mass at larger rates. Most of the dust produced ($\sim 80\%$) are silicates
(mostly olivine), with traces of cordundum and iron.

The yields scale with the metallicity in the high--mass domain, owing to the quantity
of silicon available, which increases with Z.

In the low--mass domain, where carbon dust formation occurs, the results are less
sensitive to metallicity. The overall dust produced by the $Z=10^{-3}$ models
is smaller, because the minimum mass at which HBB occurs is smaller, which
limits the range of masses experiencing carbon enrichment.

\section*{Acknowledgments}
RS acknowledges that the research leading to these results has received funding from the 
European Research Council under the European Union’s Seventh Framework Programme 
(FP/2007-2013) / ERC Grant Agreement n. 306476.
The authors are indebted to the anonymous referee, for the careful reading
of the manuscript, that helped improving the quality of this work.

\end{document}